\documentclass[a4paper,11pt,preprintnumbers,floatfix,superscriptaddress]{article}
\pdfoutput=1 
\usepackage{jheppub}
\usepackage[T1]{fontenc}
\usepackage{graphicx,dcolumn,bm,epsfig,cancel,hyperref}
\usepackage{multirow}
\usepackage{amsmath}
\usepackage{amssymb, times}
\usepackage{footnote}
\usepackage[utf8]{inputenc}
\newcommand{\comment}[1]{{}}
\usepackage{color}

\newcommand{\mha}[0]{m_{h_1}}
\newcommand{\mhb}[0]{m_{h_2}}
\newcommand{\bbaa}[0]{b\bar{b}\gamma\gamma}
\newcommand{\bbtt}[0]{b\bar{b}\tau^+\tau^-}
\newcommand{\bbww}[0]{b\bar{b}W^+W^-}

\newcommand{\bbtautau}{b\bar{b} \tau^{+} \tau^{-}}
\newcommand{\bbWW}{b\bar{b} W^{+} W^{-}}
\newcommand{\bbbb}{b\bar{b} b\bar{b}}

\newcommand{\bbjj}{b\bar{b}jj}
\newcommand{\jjaa}{jj\gamma\gamma}
\newcommand{\bbaj}{b\bar{b}\gamma j}
\newcommand{\ccaj}{c\bar{c}\gamma j}
\newcommand{\ccaa}{c\bar{c}\gamma\gamma}
\newcommand{\tth}{t\bar{t}h}
\newcommand{\zh}{Zh}
\newcommand{\bbh}{b\bar{b}h}

\newcommand{\xgb}{\texttt{XGBoost }}

\begin{document}
\title{
Resonant Di-Higgs Production at Gravitational Wave Benchmarks: A Collider Study using Machine Learning 
}
\author[a]{Alexandre Alves,}

\author[b]{Tathagata Ghosh,}

\author[c]{Huai-Ke Guo,}

\author[c,1]{and Kuver Sinha\note{Corresponding author.}}

\affiliation[a]{Departamento de Física, Universidade Federal de São Paulo, UNIFESP, Diadema, Brazil}
\affiliation[b]{Department of Physics \& Astronomy, University of Hawaii, Honolulu, HI 96822, USA}
\affiliation[c]{Department of Physics and Astronomy, University of Oklahoma, Norman, OK 73019, USA}

\emailAdd{aalves@unifesp.br}
\emailAdd{tghosh@hawaii.edu}
\emailAdd{ghk@ou.edu}
\emailAdd{kuver.sinha@ou.edu}



\abstract{
We perform a complementarity study of gravitational waves and colliders in the context of electroweak phase transitions choosing as our template the xSM model, which consists of the Standard Model augmented by a real scalar. We carefully analyze the gravitational wave signal at benchmark points compatible with a first order phase transition, taking into account subtle issues pertaining to the bubble wall velocity and the hydrodynamics of the plasma. In particular, we comment on the tension between requiring bubble wall velocities small enough to produce a net baryon number through the sphaleron process, and large enough to obtain appreciable gravitational wave production. For the most promising benchmark models, we study resonant di-Higgs production at the high-luminosity LHC  using machine learning tools: a Gaussian process algorithm to jointly search for optimum cut thresholds and tuning hyperparameters, and a boosted decision trees algorithm to discriminate signal and background. The multivariate analysis on the collider side is able either to discover or provide strong statistical evidence of the benchmark points, opening the possibility for complementary searches for electroweak phase transitions in collider and gravitational wave experiments.}



\maketitle

\section{ Introduction}

Understanding the nature of the electroweak phase transition (EWPT) is a major goal in particle physics. A first order phase transition can be obtained by introducing new physics at the electroweak scale and this new physics can be explored at the high luminosity Large Hadron Collider (HL-LHC). On the other hand, a first order phase transition  can generate gravitational waves that may be within the reach of future space-based detectors. It becomes important to understand how this complementarity plays out in concrete models - for example, can one obtain regions of parameter space where \textit{all} conditions - first order phase transition, detectable gravitational waves, and a strong enough signal at the HL-LHC - are met?

The simplest template for studying these questions is the xSM model \cite{Profumo:2007wc,Profumo:2014opa,Huang:2017jws}, which consists of the Standard Model (SM) extended by a real scalar.  We make no comments about the completion of this model in the UV, the naturalness conflicts associated with  introducing yet another scalar in addition to the Higgs, etc. Rather, our philosophy is to use the xSM as the simplest extension of the Higgs sector in which a  complementary gravitational wave and collider study can be performed.

The purpose of the current paper is to first carefully explore gravitational wave signatures associated with the EWPT, and then study resonant di-Higgs production at the HL-LHC in the same context.

The new features of our study are the following:

$(i)$ While the picture of complementarity presented above is appealing, making concrete connections from gravitational wave studies to particle physics at the electroweak scale faces many technical challenges in the calculations of electroweak baryogenesis (EWBG), EWPT and gravitational waves~\cite{Morrissey:2012db}. While we do not intend to target all these challenges in one strike, we initiate a process of making this connection more solid by presenting a careful treatment of the gravitational wave calculations. 

We address several subtle issues pertaining to the bubble wall velocity and the hydrodynamics of the plasma, in particular the tension between requiring bubble wall velocities small enough to produce a net baryon number through the sphaleron process, and large enough to obtain appreciable gravitational wave production. The velocity that enters the calculations of EWBG might not be the bubble wall velocity for plasma in the modes of deflagrations and supersonic deflagrations ahead of the bubble wall, as demonstrated by hydrodynamic analysis and simulations~\cite{Espinosa:2010hh}. This has the consequence that for a large wall velocity, a much smaller velocity for EWBG can be obtained and EWPT can be accompanied by a strong gravitational wave signal~\cite{No:2011fi}. Therefore in our analysis, we make a clear distinction between these two velocities and determine their relation from a hydrodynamic analysis of the fluid profiles. 

For our benchmark models, we compute the gravitational wave energy spectra and signal-to-noise ratio for future space-based gravitational wave experiments.

$(ii)$ On the collider side, our objective is to apply the machine learning techniques initiated in \cite{Alves:2017ued} to resonant di-Higgs production, at benchmark points that are compatible with acceptable EWPT and that hold out the most optimistic prospects from gravitational wave observations. We conduct a di-Higgs study at the HL-LHC: $pp \rightarrow h_2 \rightarrow h_1h_1 \rightarrow \bbaa$, where $h_1$ denotes the SM Higgs. We carefully incorporate all relevant backgrounds in our study. In particular, we are careful to include contributions coming from jets being misidentified as photons, as well as light flavor jets or $c$-jets being misidentified as $b$-jets. 

We utilize two recent advances in the machine learning literature for our collider study. Firstly, recent results \cite{hyperopt} show that in terms of efficiency, Bayesian hyperparameter optimization of machine learning models tends to perform better than random, grid, or manual optimization. 
We use the Python library \texttt{Hyperopt}~\cite{hyperopt} to optimize cuts on kinematic variables in our study. The second tool from the machine learning community that we apply is \texttt{XGBoost}~\cite{xgb} (eXtreme Gradient Boosted Decision Trees), which has become increasingly popular among Kaggle competitors and data scientists in industry, especially since its winning performance in the \textit{HEP meets ML} Kaggle challenge. 
Unlike a simple gradient boosting classifier, where classifiers (decision trees) are added sequentially, \xgb is able to parallelize this task, leading to superior performance.  
Both cut thresholds and Boosted Decision Trees (BDT) hyperparameters are jointly optimized for maximum collider sensitivity.
\begin{table*}[t!]
\resizebox{\textwidth}{!}{
\begin{tabular}{c|c|c|c|c|c|c|c|c|c|c|c|c|c|c|c|c|c|c|c|c}
\hline
\hline
 & \multirow{2}{*}{$\cos \theta$} & $m_{h_2}$ & $v_s$ & \multirow{2}{*}{$\lambda$} & $a_1$ & \multirow{2}{*}{$a_2$} & $b_3$ & 
\multirow{2}{*}{$b_4$} & $\lambda_{111}$ & $\lambda_{211}$ & 
\multirow{2}{*}{$\frac{\lambda_{111}}{\lambda^{\text{SM}}_{111}}$}   & $\Gamma_{h_2}^{\text{tot}}$ 
& \text{BR}($h_1 h_1$) & 
$T_c$ & $T_n$ & $v_h(T_n)$ & \multirow{2}{*}{$\alpha$} & \multirow{2}{*}{$\beta/H_n$} & $v_w$ & \multirow{2}{*}{\text{SNR}(\text{LISA})}  \\
 &  & (GeV)&  (GeV) & & (GeV) & & (GeV) & & (GeV) & (GeV) & &  (GeV) & (\%) & (GeV) & (GeV) & (GeV) & & & ($v_+=0.05$)& \\ 
\hline
 \text{BM5} & 0.984 & 455. & 47.4 & 0.179 & -708. & 4.59 & -607. & 0.85 & 47.0 & 92.8 & 1.48 &
    2.06 & 30.5 & 59.3 & 33.5 & 234. & 1.88 & 127. & 0.766 & 9133. \\
 \text{BM6} & 0.986 & 511. & 40.7 & 0.185 & -744. & 5.11 & -618. & 0.82 & 46.9 & 90.5 & 1.48 &
    2.44 & 22.8 & 62.3 & 49.7 & 217. & 0.48 & 726. & 0.345 & 20. \\
 \text{BM7} & 0.988 & 563. & 40.5 & 0.188 & -845. & 5.82 & -151. & 0.08 & 47.3 & 103.0 & 1.49 &
   2.90 & 23.2 & 57.3 & 28.4 & 237. & 3.45 & 67. & 0.861 & 6537. \\
 \text{BM8} & 0.992 & 604. & 36.4 & 0.175 & -900. & 7.48 & -424. & 0.28 & 45.3 & 120.4 & 1.43 &
    2.72 & 31.9 & 56.3 & 33.9 & 232. & 1.92 & 444. & 0.770 & 7473. \\
 \text{BM9} & 0.994 & 662. & 32.9 & 0.171 & -978. & 9.19 & -542. & 0.53 & 44.4 & 133.9 & 1.40 &
    2.84 & 35.2 & 54.6 & 34.0 & 230. & 1.97 & 141. & 0.774 & 10016. \\
 \text{BM10} & 0.993 & 714. & 29.2 & 0.186 & -941. & 8.05 & 497. & 0.38 & 45.1 & 108.3 & 1.42 &
    3.31 & 18.5 & 61.2 & 52.8 & 205. & 0.41 & 1307. & 0.274 & 0.50 \\
 \text{BM11} & 0.996 & 767. & 24.5 & 0.167 & -922. & 10.35 & 575. & 0.41 & 41.6 & 118.0 & 1.31 &
   2.59 & 26.4 & 63.3 & 58.3 & 186. & 0.29 & 2586. & 0.164 & 0.00048 \\
 \text{BM12} & 0.994 & 840. & 21.7 & 0.197 & -988. & 8.71 & 356. & 0.83 & 44.1 & 73.3 & 1.39 &
    3.98 & 6.1 & 68.9 & 67.4 & 152. & 0.13 & 10730. & 0.078 & 6.48$\times 10^{-10}$ \\
\hline
\hline
\end{tabular}
}
\caption{A subset of the benchmarks used in Ref.~\cite{Huang:2017jws}(Table.I) that can give a strongly first order EWPT as well as
satisfying all phenomenological constraints. BM1-4 are neglected for reasons explained in the text. 
$\lambda_{111}$ and $\lambda_{211}$ are
cubic couplings, given with the convention of Ref.~\cite{Profumo:2007wc,Profumo:2014opa,Huang:2017jws}: $\lambda_{111}=i \lambda_{h_1 h_1 h_1}/6$ 
and $\lambda_{112} = i \lambda_{h_1 h_1 h_2}/2$.
Parameters that are relevant for EWPT and gravitational waves are also tabulated for 
each benchmark. The last column is the signal-to-noise ratio which quantifies the
gravitational wave discovery prospect at LISA. See text for more detailed explanation.
}
\label{benchmark}
\end{table*}  

 
Our paper is structured as follows. In Section \ref{sec:model}, we introduce the xSM model and settle on the benchmarks that allow a first order phase transition. In Section \ref{sec:ewpt}, we calculate the gravitational wave energy spectra and signal-to-noise ratio for several benchmark models. In Section \ref{dihiggs}, we perform our collider analysis. We end with our Conclusions. 

\section{\label{sec:model}The model}
The model ``xSM'' constitutes one of the simplest extentions of the SM where a real scalar gauge singlet $S$ is
added to the particle content. The potential for the ``xSM'' model is defined with the convention following Ref.~\cite{Profumo:2007wc,Profumo:2014opa,Huang:2017jws}:
\begin{eqnarray}
  V(H,S) &=& -\mu^2 H^{\dagger} H + \lambda (H^{\dagger}H)^2 
  + \frac{a_1}{2} H^{\dagger} H S  \nonumber \\
  &&  + \frac{a_2}{2} H^{\dagger} H S^2 + \frac{b_2}{2} S^2 + \frac{b_3}{3} S^{3} + \frac{b_4}{4}S^4.
  \label{}
\end{eqnarray} 
Here $H^{\text{T}} = (G^+, (v + h + i G^0)/\sqrt{2})$ is the SM Higgs doublet and 
$S=v_s + s$ defines the real scalar singlet. All the parameters appearing here are real.
The minimization conditions of this potential at the vacuum $(v, v_s)$ allows one to eliminate $\mu, b_2$ by 
\begin{eqnarray}
&&  \mu^2 = \lambda v^2 + \frac{1}{2} v_s (a_1 + a_2 v_s), \nonumber \\
&&   b_2  =  - \frac{1}{4v_s} [v^2(a_1+2a_2 v_s) + 4v_s^2 (b_3+b_4 v_s)] .
\end{eqnarray}
With these substitutions, the mass matrix for $(h,s)$ is found to be:
\begin{eqnarray}
  m^2 = 
  \left(
  \begin{array}{cc}
    2 \lambda v^2                  &  \frac{1}{2} a_1 v + v_s v\ a_2 \\
    \frac{1}{2} a_1 v + v_s v a_2  & \ \ \  \ \ \    v_s(b_3+2v_s b_4)-\frac{1}{4v_s} v^2 a_1 
  \end{array} 
  \right) ,\nonumber 
  \label{}
\end{eqnarray}
which can then be diagonalized by a rotation angle $\theta$. This results in the physical 
scalars $(h_1, h_2)$ in terms of the gauge eigenstates $(h,s)$:
\begin{eqnarray}
h_1 = c_{\theta} h + s_{\theta} s, \quad \quad
h_2 =-s_{\theta} h + c_{\theta} s.  
  \label{}
\end{eqnarray}
where $h_1$ is identified as the 125 GeV Higgs scalar and further $m_{h_2}>m_{h_1}$. Consequently, three of the 
potential parameters $(\lambda, a_1, a_2)$ can be replaced by three physical parameters $m_{h_1}$, $m_{h_2}$ and $\theta$:
\begin{eqnarray}
  &&  \lambda = \frac{\mha^2 c_{\theta}^2 + \mhb^2 s_{\theta}^2}{2 v^2} ,\nonumber \\
  &&  a_1 = \frac{2 v_s}{v^2}[2 v_s^2(2 b_4 + \tilde{b}_3) - \mha^2 - \mhb^2 + c_{2\theta}(\mha^2 - \mhb^2)] , \nonumber \\
  &&  a_2 = \frac{-1}{2 v^2 v_s}[-2 v_s(\mha^2+\mhb^2 - 4 b_4 v_s^2) \nonumber \\
  && \hspace{2cm}+ (\mha^2 - \mhb^2)(2c_{2\theta} v_s - v s_{2\theta}) + 4 \tilde{b}_3 v_s^3] ,
\end{eqnarray}
where $\tilde{b}_3 \equiv b_3/v_s$. 
Then the full set of independent unknown parameters are 
\begin{equation}
\centering
v_s, \quad \quad \mhb, \quad \quad \theta, \quad \quad b_3, \quad \quad b_4 ,
\end{equation}
while keeping in mind that $v$ can be solved from the Fermi constant and $m_{h_1}=125\text{GeV}$.
With the model parameters fully specified, the cubic scalar couplings that are relevant for 
di-Higgs production are $\lambda_{h_1 h_1 h_1}$ and $\lambda_{h_2 h_1 h_1}$, given by
\begin{eqnarray}
  &&  i \lambda_{h_1 h_1 h_1} = 
  6 \Big[
  \lambda  v c_{\theta }^3 + 
\frac{1}{4} c_{\theta }^2 s_{\theta } \left(2 a_2 v_s+a_1\right)+\frac{1}{2} a_2 v c_{\theta }
   s_{\theta }^2 \nonumber \\
   && \hspace{1.6cm} +\frac{1}{3} s_{\theta }^3 \left(3 b_4 v_s+b_3\right) \Big] , \nonumber\\
 &&   i \lambda_{h_1 h_1 h_2} =  
   \frac{1}{2} \Big[-2 c_{\theta } s_{\theta }^2 \left(2 a_2 v_s+a_1-6 b_4 v_s-2 b_3\right) \nonumber \\
     && \hspace{0.8cm}    + 4 v \left(a_2-3 \lambda \right) c_{\theta }^2 s_{\theta }
		+c_{\theta }^3 \left(2 a_2 v_s+a_1\right)-2
	   a_2 v s_{\theta }^3 \Big]. \ \ \ \ \ 
\end{eqnarray}

In the absence of mixing of the scalars when $\theta = 0$, the cubic Higgs coupling 
reduces to its SM value $i \lambda_{h_1 h_1 h_1} = 3 \mha^2/v$ 
while $i \lambda_{h_1 h_1 h_2}$ vanishes. For small $\theta$ as suggested by experimental measurements, 
the following approximation is obtained for the cubic couplings through a Taylor expansion:
\begin{eqnarray}
  i \lambda_{h_1 h_1 h_1} &=&  \frac{3 \mha^2}{v} - \frac{3\theta^2}{2 v}\left[4(2 b_4+\tilde{b}_3)v_s^2 + 3\mha^2 - 4 \mhb^2\right], \nonumber \\
  i \lambda_{h_1 h_1 h_2} &=& \theta \frac{-4(2 b_4+\tilde{b}_3)v_s^2 - 2 \mha^2 + 3 \mhb^2}{v} . 
\end{eqnarray}
The gauge and Yukawa couplings of 
$h_1$ are reduced by a factor $c_{\theta}$ and the couplings of $h_2$ are $- s_{\theta}$ times 
the SM values, that is,
\begin{eqnarray}
  \lambda_{h1 X X} =  c_{\theta} \lambda^{\text{SM}}_{h1 X X}, \quad \quad
  \lambda_{h2 X X} =- s_{\theta} \lambda^{\text{SM}}_{h2 X X},
\end{eqnarray}
where $XX$ denotes $W^+W^-$, $ZZ$ and $\bar{f}f$. 

Since it modifies the Higgs couplings, the mixing angle 
is constrained by experiments to be small. Moreover, direct searches for a heavier 
SM-like Higgs by ATLAS and CMS as well as electroweak precision measurements further constrain the parameter space of
$(\theta, m_{h_2})$. Taking these phenomenological constraints into account, Ref.~\cite{Huang:2017jws} considered
12 benchmark points with $m_{h_2} \in [250,850]$ and studied the resonant di-Higgs production in the $b\bar{b}WW$
channel. Also imposed on these benchmarks is the strongly first order EWPT criterion, to be discussed in the next
section. Several of these benchmarks are reproduced in the current work for gravitational wave and di-Higgs production studies.
These are shown in Table.~\ref{benchmark}~\footnote{
These parameters and the couplings all agree with~\cite{Huang:2017jws}. Note that due to the limited precision shown in their paper, some reproduced numbers here differ slightly from their values. It should also be mentioned that 
in~\cite{Huang:2017jws}, a different parametrization is used with the parameter $a_2$ replaced by $m_{h_1}$. Therefore the independent set of parameters is $v_s, \lambda, a_1, m_{h_1}, b_3, b_4$. However in this method, for benchmarks BM1-3 generated 
in~\cite{Huang:2017jws}, the roles of $h_1$ and $h_2$ are switched, and we do not consider them further.} 

\section{Electroweak Phase Transition and Gravitational Waves\label{sec:ewpt}}

Ever since the first detection of gravitational waves from binary black hole mergers by the LIGO and Virgo
collaborations~\cite{Abbott:2016blz}, gravitational waves have become an increasingly important new tool for studying astronomy and cosmology in addition to 
testing the general relativity of gravity in the strong field regime. More importantly, future space-based interferometer 
gravitational wave detectors, such as the Laser Interferometer Space Antenna(LISA)~\cite{Audley:2017drz}, 
can probe gravitational waves at the milihertz level, which is right the 
frequency 
range of the gravitational waves resulting from a first order EWPT~\cite{Caprini:2015zlo,Cai:2017cbj,Weir:2017wfa}. Thus gravitational wave studies 
present a new window for looking into details of the mechanism of electroweak symmetry breaking, 
complementary to direct searches at colliders and precision measurements at the low energy intensity 
frontier~\cite{Huang:2016cjm,Hashino:2016rvx,Hashino:2016xoj,Beniwal:2017eik,Croon:2018erz}.
This complementarity between traditional particle physics techniques and gravitational wave detections 
can then provide a more complete picture to understanding the physical mechanism for baryon number generation and
solving the long standing baryon asymmetry problem of the universe.

\subsection{Electroweak Phase Transition}
The starting point for analyzing the EWPT is the calculation of the finite temperature effective
potential, which typically involves the inclusion of the tree level effective potential, 
the conventional one loop Coleman-Weinberg term~\cite{Coleman:1973jx},
the one loop finite temperature corrections~\cite{Quiros:1999jp} and the daisy resummation~\cite{Parwani:1991gq,Gross:1980br}. 
It is known that there is a gauge parameter dependence in the  effective potential thus calculated~\cite{Nielsen:1975fs}. However a gauge
invariant effective potential can be obtained by doing a high temperature expansion with the 
result equivalent to including only the thermal mass corrections~\cite{Patel:2011th}. Here the 
gauge invariant effective potential is found to be:
\begin{eqnarray}
 && V(h,s,T) = - \frac{1}{2} [\mu^2 - \Pi_h(T)] h^2 
  - \frac{1}{2} [-b_2 - \Pi_s(T)] s^2 \nonumber \\
  &&\hspace{0.7cm} + \frac{1}{4} \lambda h^4 + \frac{1}{4} a_1 h^2 s + \frac{1}{4} a_2 h^2 s^2 +
  \frac{b_3}{3} s^3 + \frac{b_4}{4} s^4, \quad \quad
\end{eqnarray}
with the thermal masses given by
\begin{eqnarray}
  &&  \Pi_h(T) = \left( \frac{2 m_W^2 + m_Z^2 + 2 m_t^2}{4 v^2} + \frac{\lambda}{2} + \frac{a_2}{24} \right) T^2, \nonumber \\
  &&  \Pi_s(T) = \left( \frac{a_2}{6} + \frac{b_4}{4} \right) T^2,
\end{eqnarray}
where we have written the gauge and Yukawa couplings in terms of the physical masses of $W$, $Z$ and the $t$-quark.
\comment{
It is convenient to reorganize above effective potential in the following illuminating form,
\begin{eqnarray}
  V(h,s,T) &=& D_h (T^2 - T_h^2) h^2 + D_s (T^2 - T_s^2) s^2 + \frac{1}{4} \lambda h^4\nonumber \\
  && + \frac{1}{4} a_1 h^2 s + \frac{1}{4} a_2 h^2 s^2 +
  \frac{b_3}{3} s^3 + \frac{b_4}{4} s^4, \quad \quad
\end{eqnarray}
}
In the above effective potential~\footnote{
Note that the above effective potential can also be written in  cylindrical coordinates to be 
compared with the result in Ref.~\cite{Profumo:2007wc,Profumo:2014opa,Huang:2017jws}}, 
it is the cubic terms that allow the realization of a first order EWPT by providing a tree level barrier.
This fact also greatly mitigates the possible effect due to the neglection of higher order terms
in the approach of calculating effective potential here~\cite{Profumo:2014opa}.

We further note that in the above effective potential, we have neglected a tadpole term proportional to $T^2 s$, coming from the
terms proportional to $a_1$ and $b_3$ in the tree level potential. The effect of this term has been found to be 
numerically negligible~\cite{Profumo:2007wc} as it is suppressed by $v_s/v_{\text{EW}}$.

Among the physical parameters that characterize the dynamics of a first order EWPT, the following enter the calculation of the 
gravitational waves~\cite{Caprini:2015zlo}:
\begin{equation}
\centering
T_c, \quad T_n, \quad \alpha, \quad \beta, \quad v_w .
\end{equation}
Here $T_c$ is the critical temperature at which the stable and metastable vacua become degenerate, 
$T_n$ is the nucleation temperature when a significant fraction of the space is filled with nucleated
electroweak bubbles,
$\alpha$ is the ratio between the released energy from the EWPT and the total radiation energy density at $T_n$,
$\beta$ denotes approximately the inverse time duration of the EWPT and $v_w$ is the bubble wall 
velocity~\cite{Chao:2017vrq,Bian:2017wfv,Chao:2017ilw}. 
We use \texttt{CosmoTransitions}~\cite{Wainwright:2011kj} to 
trace the evolution of the phases as temperature drops and solve the bounce solutions to determine $T_c$, $T_n$, 
$\alpha$ and $\beta$~\footnote{
Aside from BM1, BM2 and BM3 in Ref.~\cite{Huang:2017jws} which we neglected for reasons explained earlier, we 
found that for BM4, the nucleation temperature $T_n$ cannot be obtained. This may be due to the limited precision 
presented there since it is known that tunneling calculations are very sensitive to input parameters.}. 
These results are added to Table.~\ref{benchmark} for each benchmark. The following
comments are important regarding these benchmarks:
\begin{itemize}
\item 
To avoid washout of the generated baryons inside the electroweak bubbles, the strongly first order EWPT 
criterion $v_h(T_n)/T_n \gtrsim 1$~\cite{Cline:2006ts,Morrissey:2012db} needs to be met, which effectively quenches the sphaleron process inside the bubbles. All the benchmarks presented in Table.~\ref{benchmark} satisfy this condition.
\begin{figure}[t]
\centering
  \includegraphics[width=0.4\columnwidth]{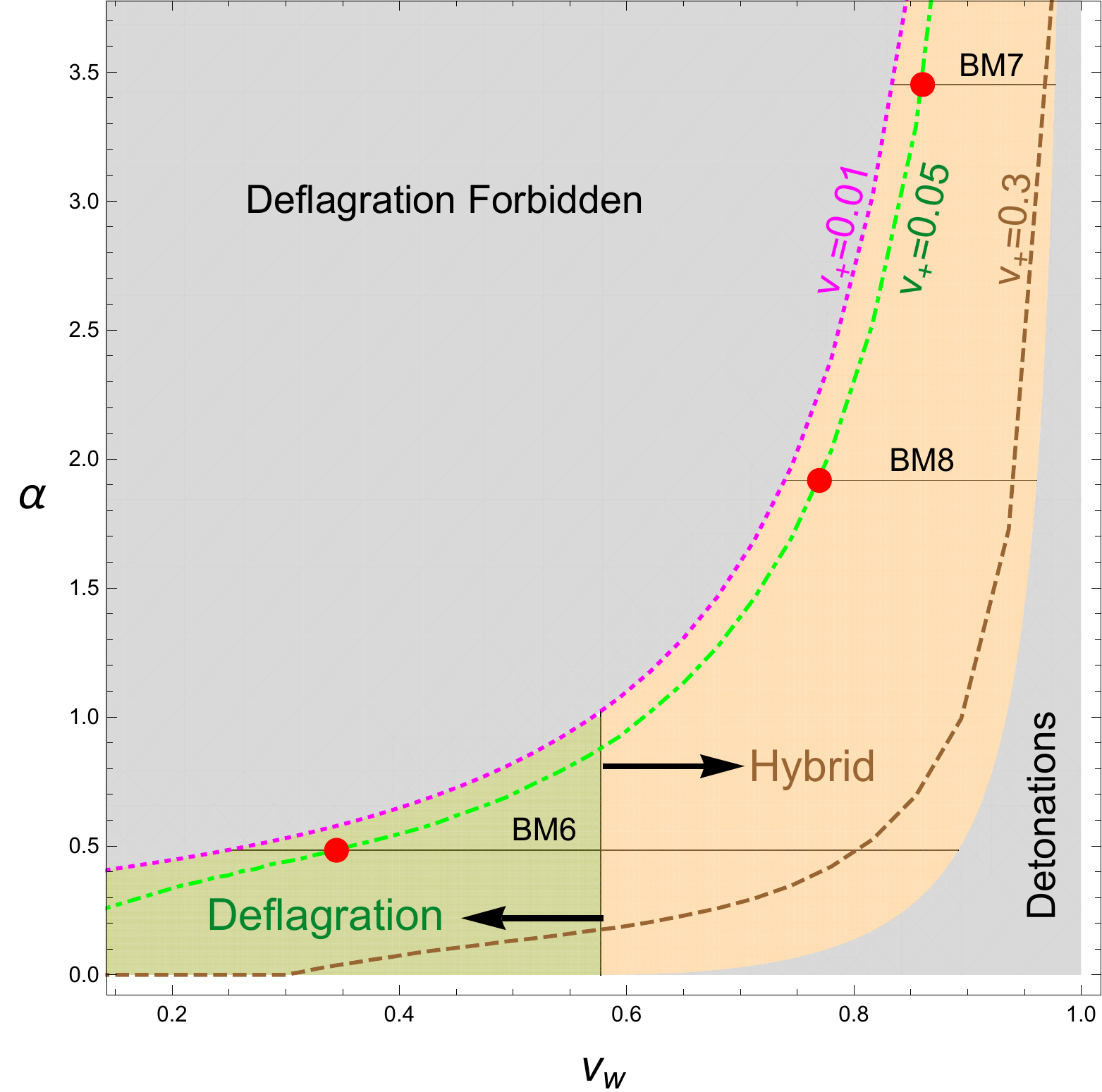}
  \includegraphics[width=0.4\columnwidth]{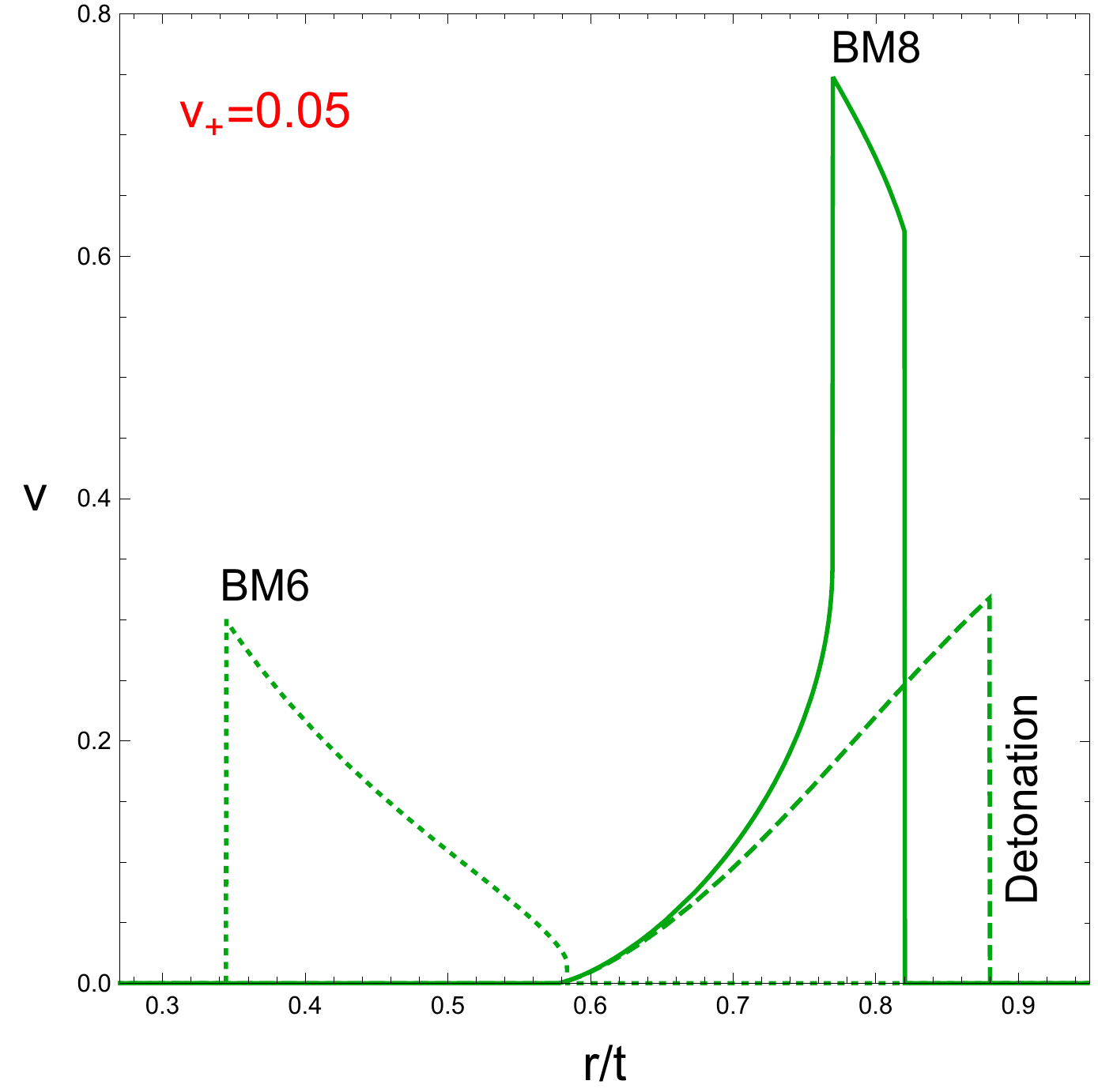}
\caption{\label{fig:plasma}
Left panel: constraint on the plane $(v_w, \alpha)$ from hydrodynamic considerations. 
Right panel: representative velocity profiles for plasma surrounding the bubble wall for each of the 
three modes with $r$ the distance from the bubble wall center and $t$ starting from the onset of the phase transition.
See text for detailed explanations.
}
\end{figure}
\item 
Currently there is large uncertaintity with the determination of the bubble wall velocity $v_w$, so it is usually
taken as a free parameter in the calculations of EWBG, EWPT and gravitational waves. It is however not 
entirely free as there are constraints from admitting consistent hydrodynamic solutions of the plasma at the time of phase transition, to be discussed in the following.
\item 
Very strong phase transitions are observed for BM5, BM7, BM8 and BM9 as their values of $\alpha$ are all larger 
than 1. A hydrodynamical analysis of the plasma surrounding the bubbles shows that the profiles of the plasma can
be classified into three categories~\cite{Espinosa:2010hh}: deflagrations, detonations and supersonic 
deflagrations (aka hybrid)~\cite{KurkiSuonio:1995pp}, depending on
the value of the bubble wall velocity $v_w$. For $v_w$ smaller than the speed of sound in the plasma ($c_s=1/\sqrt{3}$), the 
plasma takes the form of deflagrations with the following properties: (a) the plasma ahead of the phase front flows
outward with non-zero velocity; (b) the plasma inside the bubbles are static. For $c_s < \xi_J(\alpha) < v_w$ where 
$\xi_J$ as a function of $\alpha$ is the velocity corresponding to the Jouguet detonation~\cite{Steinhardt:1981ct}, a detonation profile
is obtained: (a) the plasma ahead of the wall is static; (b) the plasma inside the wall flows outward. For 
intermediate values of $v_w$ with $c_s< v_w < \xi_J(\alpha)$, a supersonic deflagration mode is obtained with the feature that both the plasma ahead of and
behind the wall flow outward. An important implication relevant for the analysis here is that there is a minumum value of
$v_w$ when $\alpha>1/3$ for deflagration and hybrid modes~\cite{Espinosa:2010hh}, where $v_w$ smaller than this value gives no consistent
solution. For benchmarks BM11 and BM12 both with $\alpha<1/3$, $v_w$ can take any value, while 
for BM5-10, there is a limited range for $v_w$. 

In the left panel of Fig.~\ref{fig:plasma}, we show on the plane of 
$(v_w, \alpha)$, the resulting ranges of 
$v_w$ for BM6, BM7 and BM8, denoted by  black horizontal lines that extend between the two gray region boundaries. We note that the value of $\alpha$ for BM10 is close to that of BM6, while the values of $\alpha$ for BM5 and BM9 are similar to BM8. We do not plot these cases to prevent the plot from being overcrowded. 
The left gray region is forbidden by the constraint mentioned above, while the right gray region gives a $v_w$ too
fast for EWBG to work~\footnote{There may also be an additional excluded region on this plane from the consideration
that for fixed $v_w$, $\alpha$ needs to be larger than a critical value to surmount a possible hydrodynamic 
obstruction~\cite{Konstandin:2010dm,No:2011fi}. This mainly affects small values of $\alpha$ and is not considered here}. 
The allowed regions in this plot are the light green region for deflagration and the brown
region for supersonic deflagration. We also show three representative fluid profiles in each of the modes
in the right panel of Fig.~\ref{fig:plasma}.

\item The usual consensus for EWBG calculations is that the bubble wall velocity needs to be 
sufficiently small to allow diffusion of particles ahead of the wall and to produce net baryon number through 
the sphaleron process, with a typical value of $v_w=0.05$ (see for example~\cite{John:2000zq,Cirigliano:2006dg,Chung:2009qs,Chao:2014dpa,Guo:2016ixx,White:2016nbo}). However such 
small velocities would weaken gravitational wave production. The story changes when the hydrodynamic properties of the 
plasma surrounding the bubble wall are taken into account, and the dilemma between successful baryon number generation and 
a strong gravitational signal may be avoided. The reason is that the plasma ahead of the wall can be stirred by the
expanding wall  
and gain a velocity in the deflagration and hybrid modes. This has the consequence that in the wall frame the 
plasma would hit the wall with a velocity 
$v_+$ that is different from $v_w$~\cite{Espinosa:2010hh,No:2011fi} and it is $v_+$ rather than $v_w$ that should
enter the calculations of EWBG. While a definitive justification of this argument would require 
analyzing the transport behavior of the particle species surrounding the wall in the above picture, we assume tentatively
that this is true in this work(see Ref.~\cite{Kozaczuk:2015owa} for a similar discussion on this point 
in the same model). 
The contours for a subsonic $v_+$ with values of $0.3$, $0.05$ and $0.01$ are shown in the left panel of 
Fig.~\ref{fig:plasma}. We can see that $v_+$ decreases as $\alpha$ increases for fixed $v_w$, with the contour 
$v_+=0$ coinciding with the boundary of the left gray region. Assuming $v_+=0.05$ is used for EWBG calculations,
we locate the value of $v_w$, which corresponds to the intersection point of this contour with the horizon line
of each benchmark, represented as a red point. The $v_w$ found in this way is used to calculate the 
gravitational wave energy spectrum. 
\end{itemize}
With above problems properly taken care of, we can now calculate the gravitational waves resulting from the EWPT.

\subsection{Gravitational Waves}

A stochastic background of gravitational waves can be generated during a first order EWPT from mainly three
sources: collisions of the electroweak bubbles~\cite{Kosowsky:1991ua,Kosowsky:1992rz,Kosowsky:1992vn,Huber:2008hg,Jinno:2016vai,Jinno:2017fby}, bulk motion of the plasma in the form of sound waves~\cite{Hindmarsh:2013xza,Hindmarsh:2015qta} and 
Magnetohydrodynamic (MHD) turbulence~\cite{Caprini:2009yp,Binetruy:2012ze}(see Ref.~\cite{Caprini:2015zlo,Cai:2017cbj,Weir:2017wfa} for recent reviews).
The total resulting energy spectrum can be written approximately as the sum of these contributions:
\begin{equation}
  \Omega_{\text{GW}}h^{2} \simeq \Omega_{\text{col}}h^{2}+\Omega_{\text{sw}}h^{2}+\Omega_{\text{turb}}h^{2} .
\end{equation}
While earlier studies of gravitational wave production from EWPT have focused on bubble collisions, 
recent advances in numerical simulations show 
that the long lasting sound waves during and after the EWPT give the dominant contribution to the gravitational wave
production~\cite{Hindmarsh:2013xza,Hindmarsh:2015qta} and the contribution from bubble collision can be 
neglected~\cite{Bodeker:2017cim}. From such numerical simulations, an analytical formula has been obtained for this kind of gravitational
wave energy spectrum~\cite{Hindmarsh:2015qta}:
\begin{eqnarray}
  &&\Omega_{\textrm{sw}}h^{2}=2.65\times10^{-6}\left( \frac{H_{\ast}}{\beta}\right) \left(\frac{\kappa_{v} \alpha}{1+\alpha} \right)^{2} 
\left( \frac{100}{g_{\ast}}\right)^{1/3} \nonumber \\
&&\hspace{1.4cm} \times v_{w} \left(\frac{f}{f_{\text{sw}}} \right)^{3} \left( \frac{7}{4+3(f/f_{\textrm{sw}})^{2}} \right) ^{7/2} \ .
\label{eq:soundwaves}
\end{eqnarray}
Here $g_{\ast}$ is the relativistic degrees of freedom in the plasma, $H_{\ast}$ is the Hubble parameter 
at $T_{\ast}$  when the phase transition has completed and has a value close to that evaluated at the nucleation
temperature $H(T_n)$ for not very long EPWT. 
We take $T_{\ast} = T_n (1+\kappa_T \alpha)^{1/4}$ where
the fraction of vacuum energy goes to heating the plasma is given by $\kappa_T \approx 1 - \kappa_v$~\cite{Espinosa:2010hh}.
Moreover, $f_{\text{sw}}$ is the present peak frequency which is the redshifted value of the peak frequency at the time of 
EWPT($=2 \beta/(\sqrt{3}v_w)$):
 \begin{equation}
f_{\textrm{sw}}=1.9\times10^{-5}\frac{1}{v_{w}}\left(\frac{\beta}{H_{\ast}} \right) \left( \frac{T_{\ast}}{100\textrm{GeV}} \right) \left( \frac{g_{\ast}}{100}\right)^{1/6} \textrm{Hz} .
\end{equation}
The factor $\kappa_{v}$ is the fraction of latent heat that is transformed into the 
bulk motion of the fluid and can be calculated as a function of ($\alpha$, $v_w$) by analyzing the energy budget 
during the EWPT~\cite{Espinosa:2010hh}. 
We note that a more recent numerical simulation by the same group~\cite{Hindmarsh:2017gnf} obtained a slightly
enhanced $\Omega_{\text{sw}} h^2$ and a slightly reduced peak frequency $f_{\text{sw}}$. 
\begin{figure}[t]
\centering
  \includegraphics[width=0.5\columnwidth]{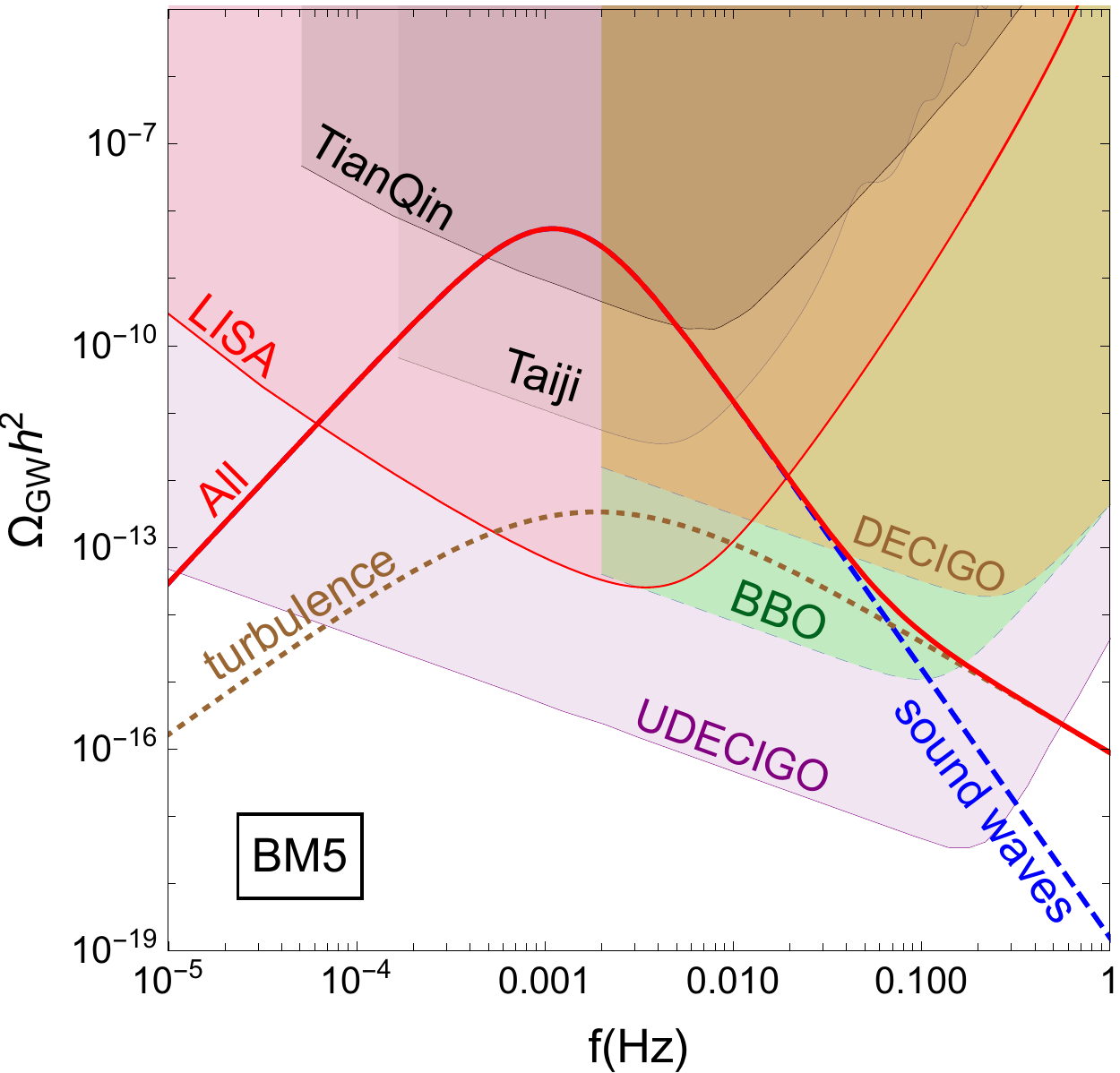}
\caption{\label{fig:gw}Gravitational wave energy spectrum for BM5 together with experimentally sensitive regions
on the top. See text for more detailed explanations of this figure. 
}
\end{figure}

It should be noted that the above numerical simulations were performed under two important assumptions, which limit
the possible applications here for some benchmarks. The first assumption is that the gravitational wave sourcing continues
at the wavenumber corresponding to the thickness of the fluid shells, which is valid when the system is linear and
requires the fluid velocity to be sufficiently smaller than unity. This is indeed what was adopted in the initial numerical
simulations~\cite{Hindmarsh:2013xza,Hindmarsh:2015qta} and in a later simulation~\cite{Hindmarsh:2017gnf} as well as in the recently 
proposed sound shell model~\cite{Hindmarsh:2016lnk}, aiming at 
understanding the origin of the shape of the gravitational wave spectra from previous simulations, which adds 
linearly the fluid velocity profiles when 
calculating the velocity power spectra. This therefore puts doubts on the effectiveness in using the above formulae for our benchmarks
with large velocities. Since there is currently no available result beyond current simulations, we assume the above
results hold for these cases and remind the reader of this possible issue here.
The second assumption is that the sourcing of gravitational waves continues until the Hubble time. This is important since
the gravitational wave energy density is directly proportional to the lifetime of the sound waves.
While there is no direct numerical simulation studies confirming this, it was found to be true in 
Ref.~\cite{Hindmarsh:2015qta,Hindmarsh:2016lnk}.

Aside from the sound waves which give the dominant gravitational wave signals, the fully ionized plasma at the time of EWPT
results in MHD turbulence, giving another source of gravitational waves.
When a possible helical component~\cite{Kahniashvili:2008pe} is neglected, the 
resulting gravitational wave energy spectrum can be modeled in a similar way~\cite{Caprini:2009yp,Binetruy:2012ze},
\begin{eqnarray}
  &&\Omega_{\textrm{turb}}h^{2}=3.35\times10^{-4}\left( \frac{H_{\ast}}{\beta}\right) \left(\frac{\kappa_{\text{turb}} 
\alpha}{1+\alpha} \right)^{3/2} \left( \frac{100}{g_{\ast}}\right)^{1/3}  \nonumber \\
  && \hspace{1.8cm}\times v_{w}  \frac{(f/f_{\textrm{turb}})^{3}}{[1+(f/f_{\textrm{turb}})]^{11/3}(1+8\pi f/h_{\ast})} ,
\label{eq:mhd}
\end{eqnarray}
where the peak frequency $f_{turb}$ corresponding to MHD is given by:
\begin{equation}
f_{\textrm{turb}}=2.7\times10^{-5}\frac{1}{v_{w}}\left(\frac{\beta}{H_{\ast}} \right) \left( \frac{T_{\ast}}{100\textrm{GeV}} \right) \left( \frac{g_{\ast}}{100}\right)^{1/6} \textrm{Hz} .
\end{equation}
Similar to $\kappa_v$, here the factor $\kappa_{\text{turb}}$ is the fraction of latent heat 
that is transferred to MHD turbulence. A recent numerical simulation shows that  
when $\kappa_{\rm turb}$ is parametrized as $\kappa_{\text{turb}}\approx \epsilon \kappa_{v} $, the numerical 
factor $\epsilon$ can vary roughly between $5 \sim 10\%$~\cite{Hindmarsh:2015qta}. 
Here we take tentatively $\epsilon = 0.1$. As has been discussed in previous section, we take the value of $v_w$ 
such that they all yield $v_+ = 0.05$, a good choice for EWBG calculations.

Adding the results given in Eq.~\ref{eq:soundwaves} and Eq.~\ref{eq:mhd}, we can then obtain the total gravitational wave energy 
density spectrum. For example, the resulting gravitational wave energy spectrum for BM5 is shown in Fig.~\ref{fig:gw}. The blue dashed line denotes the gravitational wave signal from sound waves and the brown dotted line from MHD turbulence, 
while the total contribution is shown with the solid red line. The color-shaded regions on the top are the experimentally sensitive regions
for several proposed space-based gravitational wave detectors: LISA introduced earlier, the Taiji~\cite{Gong:2014mca} and TianQin~\cite{Luo:2015ght} programs, Big Bang Observer (BBO), DECi-hertz Interferometer Gravitational wave Observatory (DECIGO) and Ultimate-DECIGO~\cite{Kudoh:2005as}~\footnote{The BBO and DECIGO data are taken from the website~\url{http://rhcole.com/apps/GWplotter/}}. 

We note that astrophysical foregrounds, such as the unresolved stochastic gravitational waves from the population of 
white dwarf binaries in the Galaxy~\cite{Klein:2015hvg}, might change the above sensitivity curves slightly. While a 
future precise modeling of these forgrounds is definitely important in discovering the stochastic gravitational 
wave of cosmological origin when the detector is online and taking data, we find it is sufficient to use above 
sensitivity curves in this study.

\begin{figure}[t]
\centering
  \includegraphics[width=0.7\columnwidth]{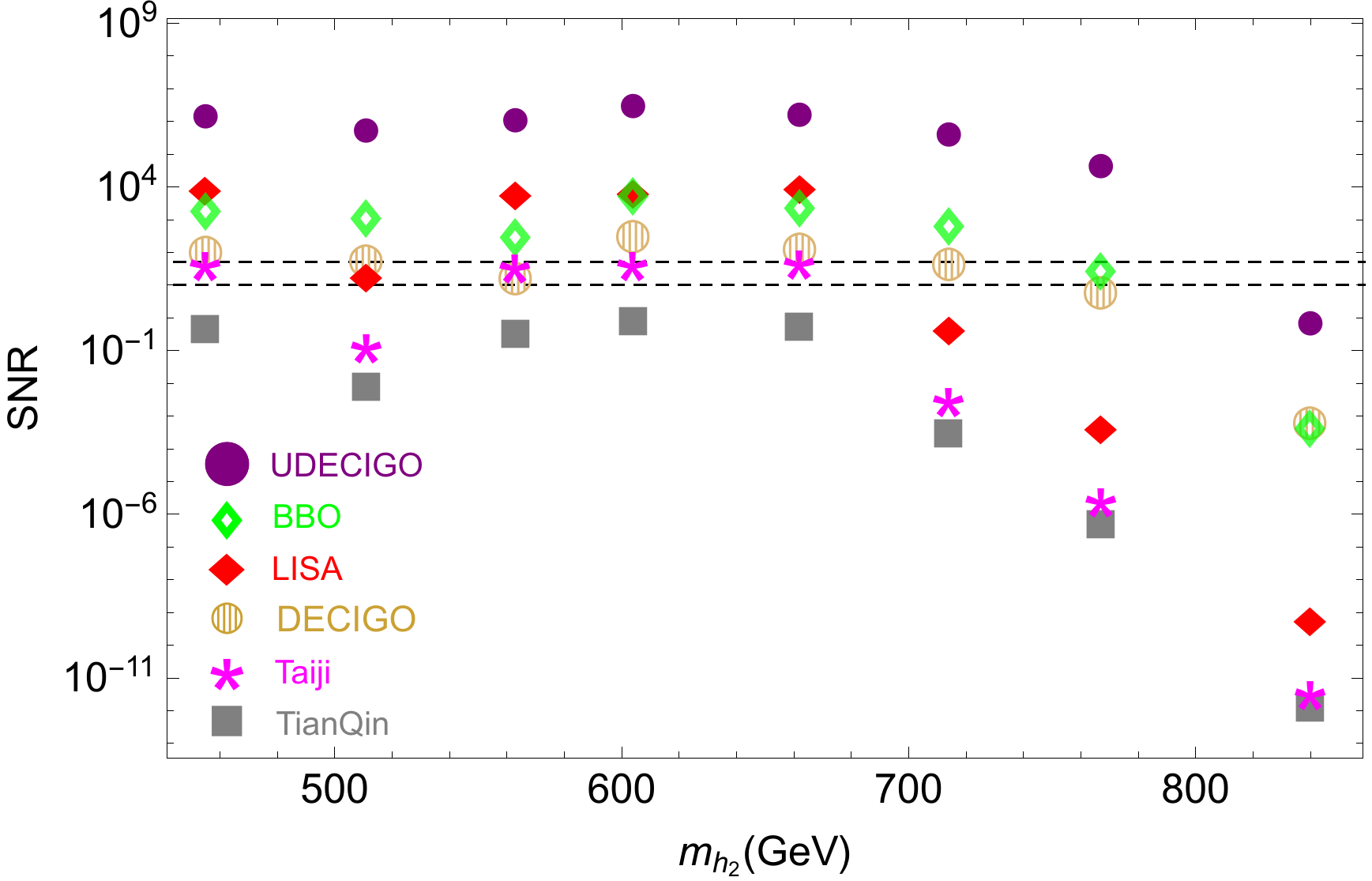}

\caption{\label{fig:snr}
The SNR of the gravitational wave signals versus $m_{h_2}$ for the benchmarks shown in Table 1 for
proposed space-based gravitational wave detectors. The two horizontal dashed lines are suggested thresholds for detection:
$\text{SNR}=10$ and $\text{SNR}=50$, depending on detector configurations.
  }
\end{figure}
To assess the discovery prospects of the generated gravitational waves, we calculate the signal-to-noise 
ratio with the definition adopted by Ref.~\cite{Caprini:2015zlo}:
\begin{eqnarray}
  \text{SNR} = \sqrt{\delta \times \mathcal{T} \int_{f_{\text{min}}}^{f_{\text{max}}} df 
    \left[
      \frac{h^2 \Omega_{\text{GW}}(f)}{h^2 \Omega_{\text{exp}}(f)} 
  \right]^2} ,
\end{eqnarray}
where $h^2 \Omega_{\text{exp}}(f)$ is the experimental sensitivity for the proposed experiments listed above and 
$\mathcal{T}$ is the mission duration in years for each experiment, assumed to be 5 here. 

The additional factor $\delta$ comes from the number of independent channels for cross-correlated detectors, which 
equals $2$ for BBO as well as UDECIGO and $1$ for the others~\cite{Thrane:2013oya}.

For the LISA configurations with
four links, the suggested threshold SNR for discovery is 50~\cite{Caprini:2015zlo}. 
For the six link configurations as drawn here, the 
uncorrelated noise reduction technique can be used and the suggested SNR threshold is 10~\cite{Caprini:2015zlo}. 
We show the SNR for the benchmarks 
versus $m_{h_2}$ in Fig.~\ref{fig:snr}. The SNR for LISA are also added in Table.~\ref{benchmark} for each of the benchmarks,
where it shows that BM5, BM6, BM7, BM8, BM9 all have SNR larger than $10$. In particular the SNR for BM5, BM7, BM8, 
BM9 are all much larger than 10 and for each of these cases a very strong gravitational wave signal is expected.
The last three benchmarks BM10-12 give gravitational wave signals too weak to be detected by LISA, Taiji and TianQin
but some may be detected by other proposed detectors.

\section{Di-Higgs Analysis} \label{dihiggs}

Probing double Higgs production is a major goal of the HL-LHC \cite{Koffas:2017cam, Davey:2017obx, Morse:2017efg, Goncalves:2018qas, Kim:2018uty,Kim:2018cxf}. Many theoretical studies of double Higgs production within the Standard Model have been conducted, for example in final states like $\bbaa$ ~\cite{Baur:2003gp, Baglio:2012np, Huang:2015tdv, Azatov:2015oxa, Chang:2018uwu, Ellis:2018gqa}, $\bbtautau$ \cite{Baur:2003gpa, Dolan:2012rv}, $\bbWW$ \cite{Papaefstathiou:2012qe}, and $\bbbb$ \cite{deLima:2014dta, Behr:2015oqq}. Moreover, resonant di-Higgs production has also been studied by various authors \cite{CerdaAlberich:2018tkw, Reichert:2017puo, Huang:2017jws,Adhikary:2017jtu,Chen:2014ask, Lewis:2017dme, Chen:2017qcz} in the context of EWBG~\cite{Morrissey:2012db}.

\begin{figure}[h]
\centering
  \includegraphics[width=0.3\columnwidth]{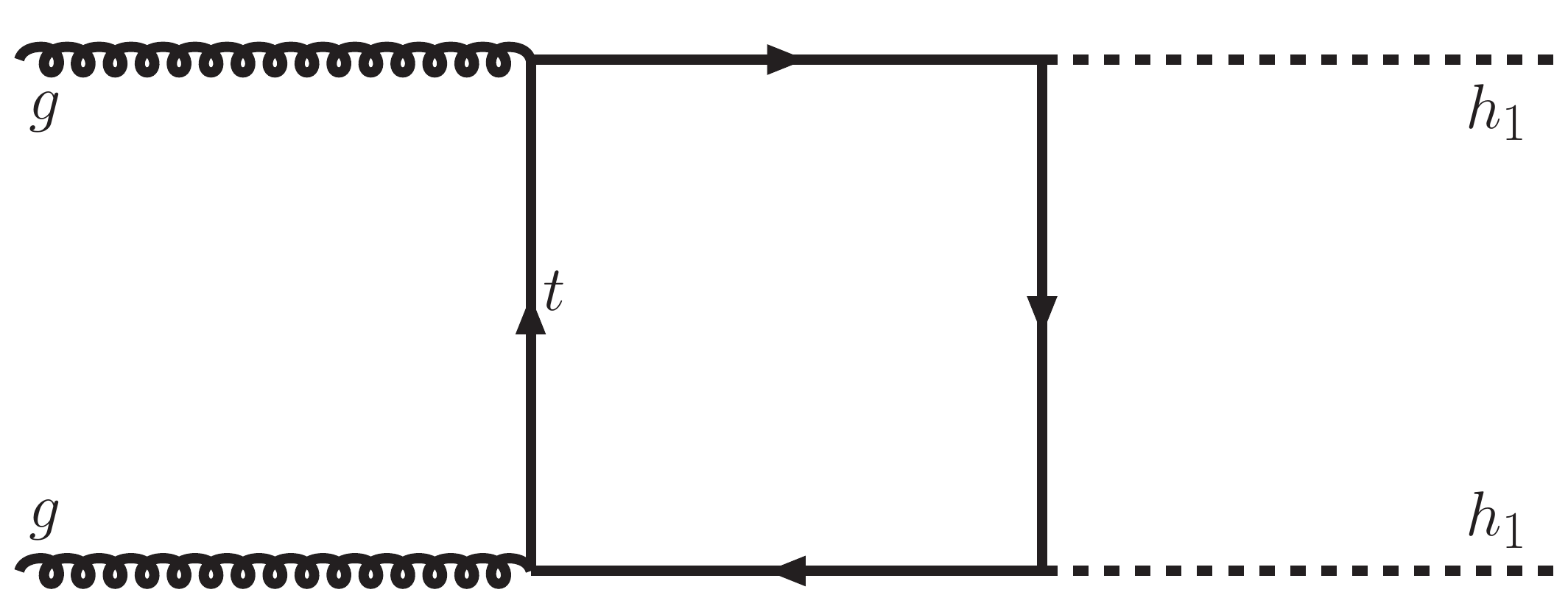}
  \includegraphics[width=0.3\columnwidth]{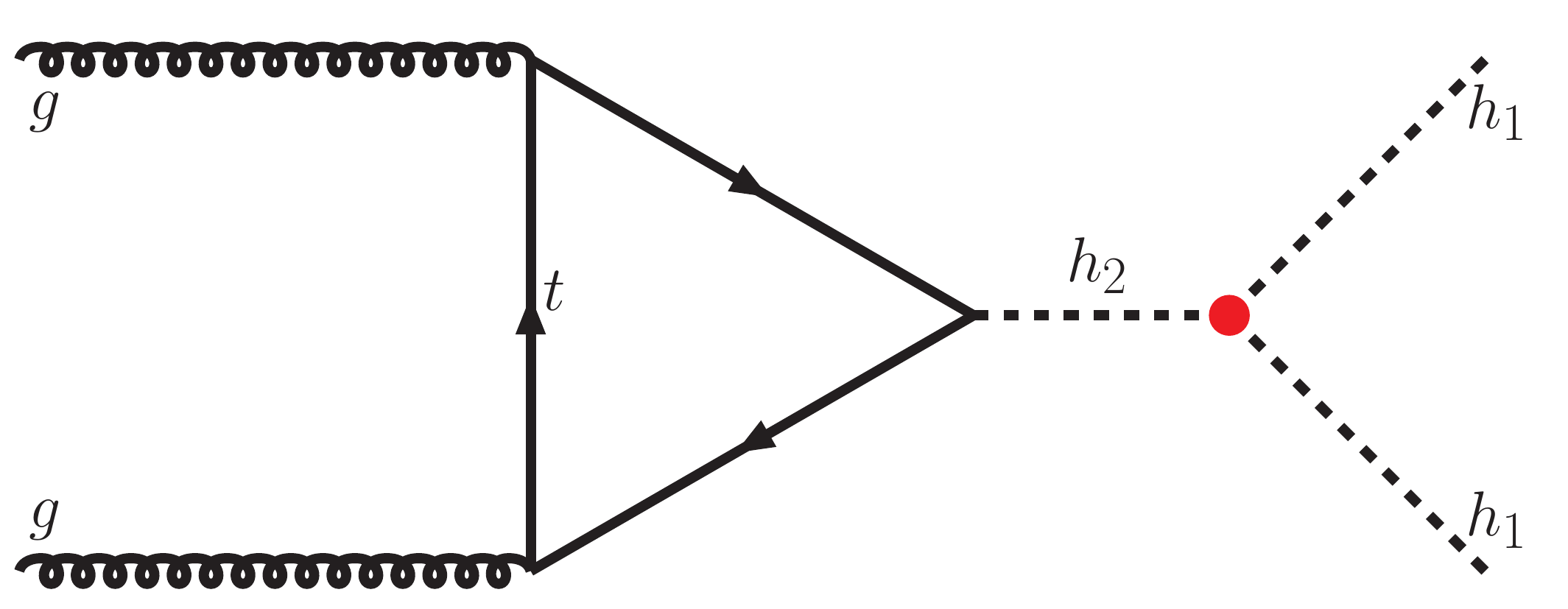}
  \includegraphics[width=0.3\columnwidth]{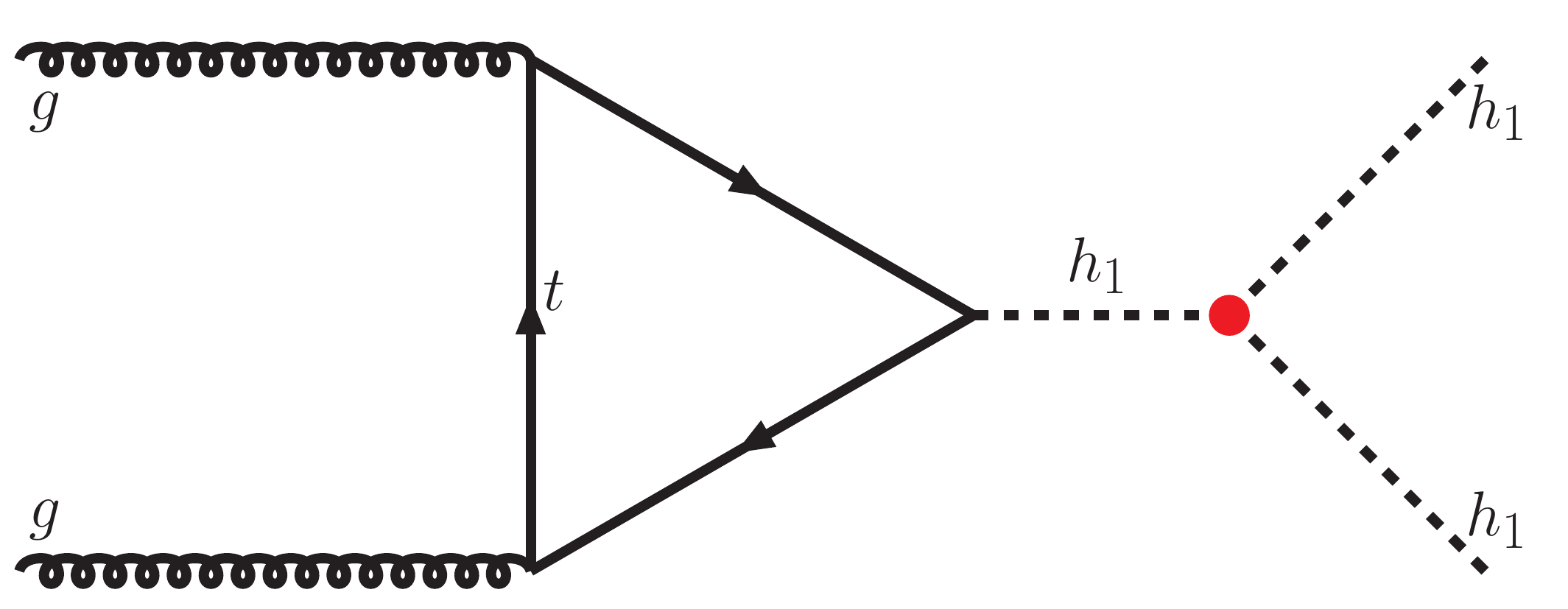}
  \caption{Representative Feynman diagrams for Di-Higgs production. \label{fig:feynman}}
\end{figure}

In this Section, we study the collider prospects of probing the benchmark points for which a large SNR for proposed gravitational wave detectors has been calculated in the previous Section. The xSM model predicts a resonant di-Higgs production $pp \rightarrow h_2 \rightarrow h_1h_1 \rightarrow \bbaa$ which is the channel that we will explore. Double Higgs production occurs through the three contributions depicted in Fig.~\ref{fig:feynman}. The non-resonant component involves the box diagram and the diagram with the trilinear Higgs coupling, while the resonant contribution corresponds to the diagram with $h_2$ in the $s$-channel. 

The non-resonant production cross section is strongly dependent on the size of $\lambda$, with a minimum at $\sim 0.31$ due to destructive interference between the box and the triangle diagrams. The benchmark points considered in this work all exhibit values of $\lambda$ between the SM value of $0.13$ and $0.2$, and for these points the non-resonant production cross section is suppressed compared to the SM. This suppression is partly compensated by the resonant contribution. We checked that the interference between the resonant and non-resonant contributions is negligible, so the contributions can be added incoherently. 

\begin{figure}[t]
\centering
  \includegraphics[width=0.65\columnwidth]{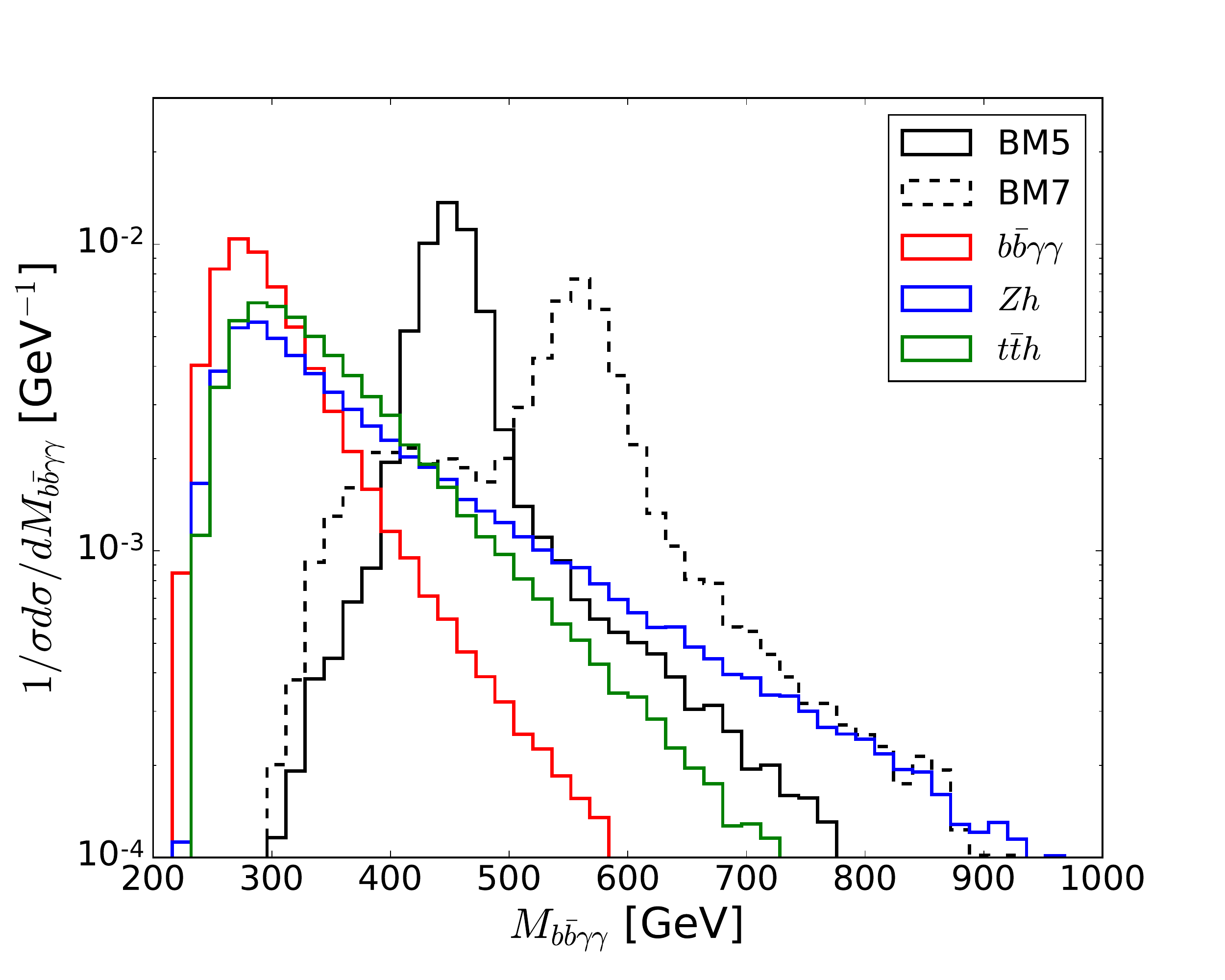}
\caption{\label{fig:mhh} The $b\bar{b}\gamma\gamma$ invariant mass distribution for signals, BM5 (solid black) and BM7 (dashed black) benchmark points, and the main backgrounds.}
\end{figure}

While the resonant di-Higgs production cross section drops rapidly as the mass of $h_2$ is increased, the resonance peak of the $h_1h_1$ invariant mass becomes easier to identify in the tail of the background distribution as shown in Fig.~\ref{fig:mhh}. Taking this tradeoff into account, and noticing that BM5 and BM7 provide acceptable SNR in the gravitational waves calculation, we take these two benchmarks as the most promising ones to be probed at the HL-LHC.

We study the $\bbaa$ channel, which is currently the most promising channel to study the double Higgs production in the SM~\cite{Baur:2003gp, Baglio:2012np, Huang:2015tdv, Azatov:2015oxa, Kim:2018uty, Barger:2013jfa, Dawson:2013bba, Alves:2017ued}. Recently, the fully leptonic $\bbww$ channel was studied in the context of the xSM~\cite{Huang:2017jws}. This channel presents better prospects than $\bbtt$ and $\bbaa$ for scalar masses greater than around 450 GeV. However, the signal-to-background ratio for the BM5 and BM7 points is $\sim 0.1$ which may be an issue if the systematic uncertainties in $t\bar{t}$ backgrounds are not very well controlled. Moreover, the presence of two neutrinos precludes the reconstruction of the scalar resonance. The $\bbaa$ channel, on the other hand, is cleaner and permits the reconstruction of the Higgses, while its cross section is much smaller than the $\bbww$ and $\bbtt$ channels. 

In Ref.~\cite{Alves:2017ued}, we found that the challenge of controlling the systematic uncertainties can be addressed by judiciously adjusting the selection criteria in order to raise the signal-to-background ratio. A full comparative study across different channels using our methods would be interesting, and is left for future study.

 Inclusive di-Higgs production was simulated with \texttt{MadGraph5\_aMC}~\cite{Alwall:2014hca} at $\sqrt{s}=14$ TeV and NN23LO1 PDFs~\cite{Ball:2013hta}. We multiply the non-resonant LO rates by the NNLO QCD K-factor of 2.27~\cite{deFlorian:2013uza}, the resonant one by the NNLL QCD K-factor of 2.5~\cite{Catani:2003zt} and add them together to get the total cross section. This is justifiable once the contributions do not interfere. Besides the fact that the K-factors for the two contributions are similar, the kinematic cuts enhance the resonant contribution to eliminate backgrounds more efficiently. The total di-Higgs production cross section is thus approximated as described, and  our signal events are weighted accordingly. 


The signal cross sections are displayed in Table~\ref{tab:xsec}. The Higgs bosons are decayed into bottom quarks and photons with the {\tt MadSpin} module of {\tt MadGraph5}. 
 We  pass our  simulated  events  to \texttt{Pythia8}~\cite{Sjostrand:2014zea} for hadronization and showering of jets.
\texttt{FastJet}~\cite{Cacciari:2011ma} is employed for clustering of jets and \texttt{Delphes}~\cite{deFavereau:2013fsa} for detector effects.

 The backgrounds were also simulated within the same framework~\footnote{The relevant backgrounds which contain a Higgs in the final state, the Higgs boson has been decayed within {\tt Pythia8}.} and their total yield is shown in Table~\ref{tab:xsec}. The backgrounds accounted for include  $\bbaa$, $\zh$ ($Z\to b\bar{b}$ and $h\to\gamma\gamma$), $\bbh$ ($h\to\gamma\gamma$), $\tth \rightarrow b\bar{b}+\gamma\gamma+X$, $\jjaa$ (the light-jets $jj$ are mistaken for $b$-jets), $\bbjj$ (the light-jets $jj$ are mistaken for photons), $\ccaa$ (a $c$-jet is mistagged as a $b$-jet), $\bbaj$ (the light-jet is mistaken for a photon), and $\ccaj$ (the $c$-jets are mistagged as $b$-jets and the light-jet as a photon), nine in total. 
 
 The first four backgrounds are generated with one extra parton radiation to better simulate the kinematic distributions, and MLM scheme~\cite{Mangano:2006rw} of jet-parton matching is used to avoid double counting. Their cross section normalizations were taken from Ref.~\cite{Azatov:2015oxa}. 
All the other five backgrounds are normalized by their NLO QCD rates from \cite{Alwall:2014hca} but their simulation do not involve extra jets. The probability of a light-jet to be mistagged as a photon is taken to be $1.2\times 10^{-4}$, although this may be an underestimate if pileup is taken into account.
 
 We note that several previous studies underestimated the background, and/or did not take into account light flavor jets or $c-$jets being misidentified as $b$-jets, or jets being misidentified as photons. We correctly take into account $\bbaj$, $\ccaa$ and $\ccaj$ backgrounds in our work. We  assume  a 70\% $b$-tagging efficiency for jet $p_T>100$ GeV, a photon efficiency of 90\% and a 20(5)\% mistagging factor for $c$($j$)-jets. We refer to \cite{Alves:2017ued} for further details of the background simulation and normalization. 

The basic event selection requirements are two b-tagged jets with $p_T(b)>30$ GeV, and two photons with $p_T(\gamma)>20$ GeV, all within $|\eta|<2.5$.  Bottom jets and photons pairs are further required to reconstruct a 125 GeV Higgs boson with $|M_{bb(\gamma\gamma)}-125|<25$ GeV, and all identified particles are isolated from any other reconstructed object within a cone of $\Delta R=0.4$ around the particle's 3-momentum.

In order to improve the statistical significance of the signal hypothesis against the background hypothesis, we used machine learning tools. First, we used an algorithm to learn the best kinematic cut thresholds in order to maximize the significance metric. This algorithm was shown to increase the significance of the non-resonant di-Higgs study in the $\bbaa$ channel up to 50\% without relying on any other multivariate analysis~\cite{Alves:2017ued}. It is based on a Gaussian process algorithm built upon the backend program \texttt{Hyperopt}~\cite{hyperopt}. We refer to~\cite{Alves:2017ued} for a detailed description of the algorithm and its usefulness in increasing the signal significance. Many other multivariate tools can be used to improve the classification of collision events as, for example, those employed in Refs.~\cite{Aaltonen:2010jr,Baldi:2014kfa,Alves:2016htj,Alves:2015dya}.

The kinematic variables chosen are: (1) the transverse momentum of the two leading bottom jets and the two leading photons, (2) the $\gamma\gamma$ invariant mass, (3) $\Delta R(\gamma,\gamma)$, the distance between the two leading photons, and (4) the $\bbaa$ invariant mass, totaling seven kinematic variables. The peak in the $\bbaa$ mass is helpful in isolating the signal events, and, in contrast to the standard analysis of non-resonant SM double Higgs production in this channel, makes the search efficient without many more variables. One interesting kinematic feature helps to explain the larger efficiency of the BM5 point. The heavy Higgs mass is right on the bulk of the non-resonant $h_1 h_1 \rightarrow \bbaa$ invariant mass after the basic selections, around 450 GeV, but for the BM7 point, it is displaced to 563 GeV as we can see in Fig.~\ref{fig:mhh}. Requiring a cut around the mass peak thus retains more non-resonant di-Higgs events for the BM5 point, raising its cut efficiency compared to BM7.
\begin{table}[t]
\centering
\begin{tabular}{c|c|c|c}
\hline\hline
 & BM5 & BM7 & Total Backgrounds \\ 
\hline
$\sigma$(fb) & 0.012 & $5.8\times 10^{-3}$  & 0.83 \\
\hline
$\varepsilon_{eff}$ & 0.4 & 0.27 & $5.4(1.5)\times 10^{-3}$ \\
\hline
$\sigma\cdot \varepsilon_{eff}\cdot L$ & 14.4 & 4.7 & 13.5(3.7) \\
\hline\hline
\end{tabular}
\caption{The signal benchmarks, BM5 and BM7 are displayed at the first two columns and the total background is  displayed in the last column. In the first row, we show the cross sections, in fb, after the basic selection discussed in the text.The second and third rows show the  cut efficiencies and the number of events after optimization assuming 3 ab$^{-1}$. The numbers in parenthesis in the last column represents the backgrounds for the cuts that maximize the BM7 point.}
\label{tab:xsec}
\end{table}

The cuts that maximize the signal significance for BM5 and BM7 benchmark points, assuming 3 ab$^{-1}$ of integrated luminosity and 10\% systematic error in the total background rates, are the following
\begin{eqnarray}
& \hbox{\underline{BM5} :} & p_T(b) > 47(30)\;\hbox{GeV},\; p_T(\gamma) > 86(49)\;\hbox{GeV}, \nonumber \\
& & \Delta R_{\gamma\gamma} < 4.4,\; |M_{\gamma\gamma}-125|<5\; \hbox{GeV}, \nonumber \\
& & |M_{\bbaa}-455| <  38\;\hbox{GeV} \\
& \hbox{\underline{BM7} :} & p_T(b) > 54(30)\;\hbox{GeV},\;\; p_T(\gamma) > 104(40),\;\hbox{GeV}, \nonumber \\
& & \Delta R_{\gamma\gamma} < 3.3,\; |M_{\gamma\gamma}-125|<5\;\hbox{GeV},\nonumber \\
& & |M_{\bbaa}-563| <  46\;\hbox{GeV.}
\end{eqnarray}

The cut selections for other systematics are similar. Because BM7 has a smaller production rate, the cuts learned by the algorithm were harder than the BM5 case in order to raise the significance. The cut efficiencies for the signals are almost three orders of magnitude larger than the backgrounds, as we can see in Table~\ref{tab:xsec}, reaching a signal to background ratio slightly larger than 1 for both signal points. 

The signal significance, assuming a 10\% systematic error in the total background rate is $3.2\sigma$ for the BM5, and $1.8\sigma$ for BM7, respectively, as shown in the third column of Table~\ref{tab:results} where results for 5\% and 15\% systematics are also shown. Note that $S/B$, displayed in the fourth column of Table~\ref{tab:results}, increases to soften the degradation of significance with the systematics in BM5 and is kept constant for BM7. This is the job of the cut optimization program~\cite{cutoptimize}. A public code of the algorithm used in this work to learn the cuts and run our multivariate analysis in an automatized way will be released in the future~\cite{cutoptimize}. 

Further improvement of the study was achieved by training boosted decision trees with \\ \texttt{XGBoost}~\cite{xgb} using the full representation of the events which comprise 28 kinematic variables: the transverse momentum of the two leading bottom jets  $p_T(bb)$ and two leading photons $p_T(\gamma \gamma)$, the $\Delta R$ distance, the invariant masses and the Barr variable~\cite{Barr:2005dz,Alves:2007xt} of all combinations of two particles, the $\bbaa$ mass, the azimuthal angle between the leptons and the bottoms pairs $\Delta\phi(bb,\ell\ell)$, and the missing transverse energy of the event. In order to tag $t\bar{t}h$ events we also used the number of leptons of the event. We averaged the results of a 10-fold cross validation to assess the robustness of our BDT training procedures. In order to obtain the best result possible, we tuned the BDT hyperparameters and the cut thresholds jointly. By doing this, we find the best compromise between cut-and-count and the multivariate analysis. 

The BDT classification increases the signal significance for both benchmark scenarios, predicting discovery for the BM5 and evidence for BM7 for systematics ranging from 5 to 15\%. In the case of  BM5, a $5\sigma$ discovery might be possible for around 2 ab$^{-1}$ in the $\bbaa$ channel.

\begin{figure}[t!]
\centering
  \includegraphics[width=0.6\columnwidth]{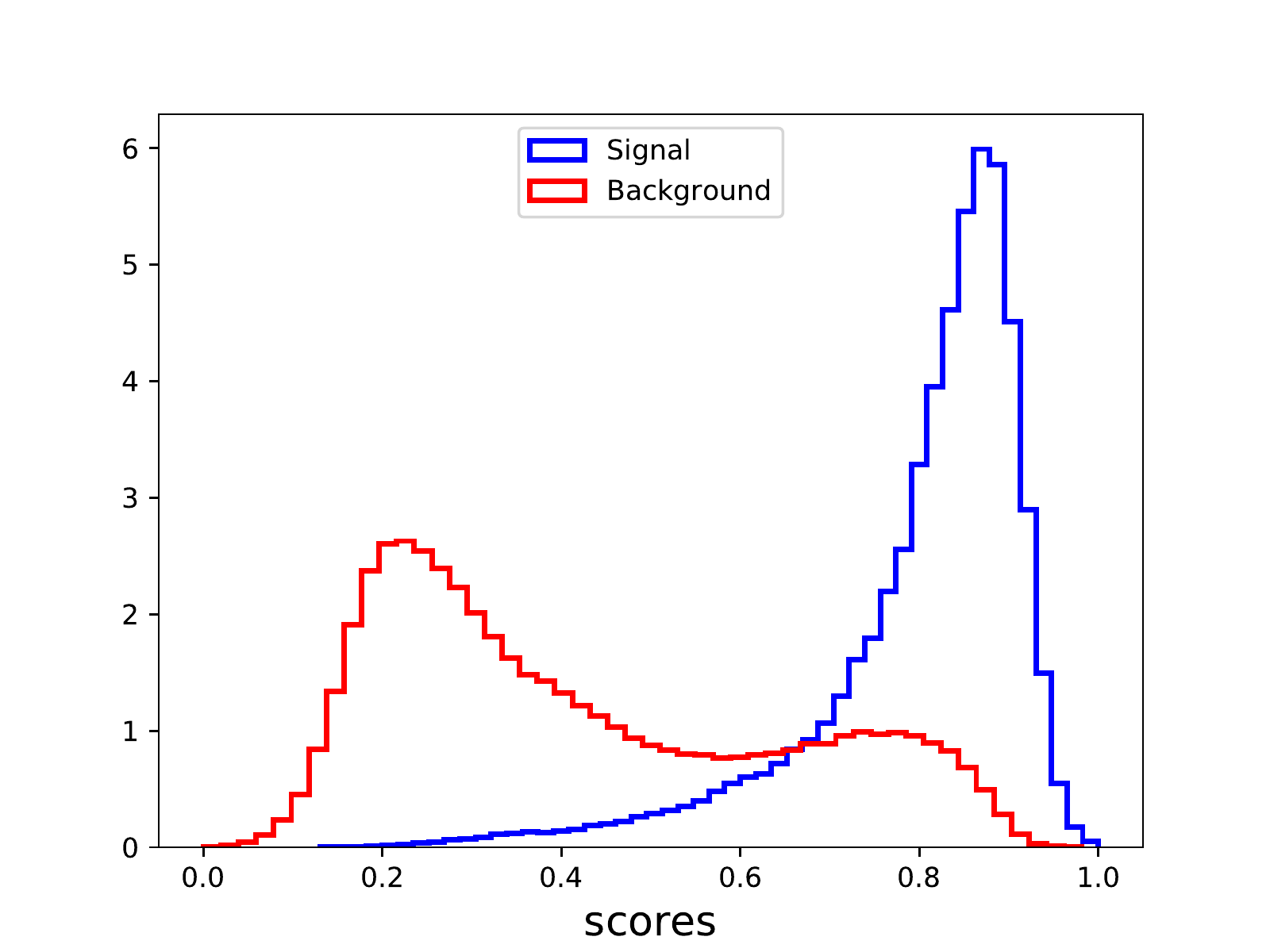}
\caption{\label{fig:bdt} The normalized BDT output scores distribution for signal(BM5) and background events after joint cuts and hyperparameters optimization. A final selection is obtained by an additional cut on these distributions.}
\end{figure}

In the last column of Table~\ref{tab:results} we display two significances: one by cutting on the BDT scores distributions of signal and background after tuning only the kinematic cuts but keeping BDT hyperparameters fixed, reaching $6.3\sigma$ and $3.3\sigma$ for BM5 and BM7, respectively, in the 5\% systematics scenario. The other number, in parenthesis, represents the significance achieved by jointly optimizing cuts and BDT hyperparameters. In this case, the significances increase slightly for all systematics but the joint optimization algorithm learns to soften $S/B$ even further, making the significance prospects insensitive to systematic uncertainties in the background rates. The final cut on the BDT scores shown in Fig.~\ref{fig:bdt} is also optimized in order to get the maximum significance possible. The typical best BDT score cut is around 0.7 which corresponds approximately to a 80\% efficiency for signals and 80\% rejection for backgrounds resulting in around 12(4) signal events against 3(1) expected background events for the BM5(BM7) point assuming 3 ab$^{-1}$. We use the profile likelihood formula of Ref.~\cite{LiMa} which approximates well the true Poissonian statistics and embodies systematic uncertainties in the background rates to compute our signal significances. 
\begin{table}[t]
\centering
\begin{tabular}{c|c|c|c|c}
\hline\hline
BM point & $\varepsilon_{sys}$(\%) & optimized cuts($\sigma$) & S/B & BDT($\sigma$) \\ 
\hline
 & 5 & 3.4 & 0.9 & 6.3(6.4)  \\
BM5  & 10 & 3.2 & 1.2 & 6.1(6.4) \\
 & 15 & 2.9 & 1.4 & 5.9(6.4) \\
\hline
& 5 & 1.9 & 0.8 & 3.3(3.4)   \\
BM7  & 10 & 1.8 & 0.8 & 3.2(3.4)  \\
 & 15 & 1.7 & 0.8 & 3.1(3.4)  \\
\hline\hline
\end{tabular}
\caption{The signal significance and the signal-to-background ratio of BM5 and BM7 benchmark points for three systematic uncertainties in the background total rates scenarios -- 5, 10 and 15\%. The results of the optimized cut-and-count and the corresponding $S/B$ achieved are displayed in the third and fourth columns, and the BDT analysis in the last column. Also, in the last column, we show in parenthesis the results for the joint BDT+cuts optimization.}
\label{tab:results}
\end{table}
%
\section{Conclusions}

Understanding the EWPT is an important goal of current and future experiments. We have explored the complementarity of the HL-LHC and proposed space-based gravitational wave detectors in achieving this goal. 

We have taken the simplest template where this complementarity can be probed - the xSM model - and studied several benchmarks that are compatible with a first order EWPT. We first calculated their gravitational wave energy spectra and signal-to-noise ratio for proposed experiments, being careful about subtle issues pertaining to the bubble wall velocity and the hydrodynamics of the plasma. Then, we took the most optimistic benchmarks and performed a collider study of double Higgs production using machine learning tools for two learning tasks: (1) to search for optimum cut thresholds and BDT hyperparameters, and (2) discriminate signal and background events with BDTs. Our results show that  state-of-the-art machine learning tools can be quite powerful in probing these processes, even assuming substantial systematic uncertainties.

There are several future directions. The tension between requiring bubble wall velocities small enough to produce a net baryon number through the sphaleron process, and large enough to obtain appreciable gravitational wave production, merits further study and a more comprehensive understanding of the parameter space in concrete models. A deeper understanding of the mechanism of gravitational wave production will be needed to obtain more realistic benchmark models. On the collider side, other final states of di-Higgs, such as $\bbww$, can be studied at these realistic benchmarks using the multivariate tools we have discussed.

\section{Acknowledgments}
A. Alves thanks Conselho Nacional de Desenvolvimento Cient\'{\i}fico (CNPq) for
its financial support, grant 307265/2017-0. K. Sinha is supported by the U.~S.~Department of Energy grant de-sc0009956.  T. Ghosh is supported by U.~S.~Department of Energy grant de-sc0010504 and in part by U.~S. National Science Foundation grant PHY-125057. H. Guo would like to thank Hao-Lin Li for helpful discussions.

\bibliographystyle{utphys}
\bibliography{mybib}

\providecommand{\href}[2]{#2}\begingroup\raggedright\begin{thebibliography}{100}

\bibitem{Profumo:2007wc}
S.~Profumo, M.~J. Ramsey-Musolf, and G.~Shaughnessy, ``{Singlet Higgs
  phenomenology and the electroweak phase transition},''
  \href{http://dx.doi.org/10.1088/1126-6708/2007/08/010}{{\em JHEP} {\bfseries
  08} (2007) 010},
\href{http://arxiv.org/abs/0705.2425}{{\ttfamily arXiv:0705.2425 [hep-ph]}}.

\bibitem{Profumo:2014opa}
S.~Profumo, M.~J. Ramsey-Musolf, C.~L. Wainwright, and P.~Winslow,
  ``{Singlet-catalyzed electroweak phase transitions and precision Higgs boson
  studies},'' \href{http://dx.doi.org/10.1103/PhysRevD.91.035018}{{\em Phys.
  Rev.} {\bfseries D91} no.~3, (2015) 035018},
\href{http://arxiv.org/abs/1407.5342}{{\ttfamily arXiv:1407.5342 [hep-ph]}}.

\bibitem{Huang:2017jws}
T.~Huang, J.~M. No, L.~Pernié, M.~Ramsey-Musolf, A.~Safonov, M.~Spannowsky,
  and P.~Winslow, ``{Resonant di-Higgs boson production in the $b{\bar b}WW$
  channel: Probing the electroweak phase transition at the LHC},''
  \href{http://dx.doi.org/10.1103/PhysRevD.96.035007}{{\em Phys. Rev.}
  {\bfseries D96} no.~3, (2017) 035007},
\href{http://arxiv.org/abs/1701.04442}{{\ttfamily arXiv:1701.04442 [hep-ph]}}.

\bibitem{Morrissey:2012db}
D.~E. Morrissey and M.~J. Ramsey-Musolf, ``{Electroweak baryogenesis},''
  \href{http://dx.doi.org/10.1088/1367-2630/14/12/125003}{{\em New J. Phys.}
  {\bfseries 14} (2012) 125003},
\href{http://arxiv.org/abs/1206.2942}{{\ttfamily arXiv:1206.2942 [hep-ph]}}.

\bibitem{Espinosa:2010hh}
J.~R. Espinosa, T.~Konstandin, J.~M. No, and G.~Servant, ``{Energy Budget of
  Cosmological First-order Phase Transitions},''
  \href{http://dx.doi.org/10.1088/1475-7516/2010/06/028}{{\em JCAP} {\bfseries
  1006} (2010) 028},
\href{http://arxiv.org/abs/1004.4187}{{\ttfamily arXiv:1004.4187 [hep-ph]}}.

\bibitem{No:2011fi}
J.~M. No, ``{Large Gravitational Wave Background Signals in Electroweak
  Baryogenesis Scenarios},''
  \href{http://dx.doi.org/10.1103/PhysRevD.84.124025}{{\em Phys. Rev.}
  {\bfseries D84} (2011) 124025},
\href{http://arxiv.org/abs/1103.2159}{{\ttfamily arXiv:1103.2159 [hep-ph]}}.

\bibitem{Alves:2017ued}
A.~Alves, T.~Ghosh, and K.~Sinha, ``{Can We Discover Double Higgs Production at
  the LHC?},'' \href{http://dx.doi.org/10.1103/PhysRevD.96.035022}{{\em Phys.
  Rev.} {\bfseries D96} no.~3, (2017) 035022},
\href{http://arxiv.org/abs/1704.07395}{{\ttfamily arXiv:1704.07395 [hep-ph]}}.

\bibitem{hyperopt}
J.~Bergstra, ``Hyperopt: Distributed asynchronous hyper-parameter
  optimization.''
\newblock \url{https://github.com/jaberg/hyperopt}.

\bibitem{xgb}
T.~Chen and C.~Guestrin, ``{XGBoost: A Scalable Tree Boosting System}.''
\newblock \url{https://github.com/dmlc/xgboost}.

\bibitem{Abbott:2016blz}
{\bfseries Virgo, LIGO Scientific} Collaboration, B.~P. Abbott {\em et~al.},
  ``{Observation of Gravitational Waves from a Binary Black Hole Merger},''
  \href{http://dx.doi.org/10.1103/PhysRevLett.116.061102}{{\em Phys. Rev.
  Lett.} {\bfseries 116} no.~6, (2016) 061102},
\href{http://arxiv.org/abs/1602.03837}{{\ttfamily arXiv:1602.03837 [gr-qc]}}.

\bibitem{Audley:2017drz}
H.~Audley {\em et~al.}, ``{Laser Interferometer Space Antenna},''
\href{http://arxiv.org/abs/1702.00786}{{\ttfamily arXiv:1702.00786
  [astro-ph.IM]}}.

\bibitem{Caprini:2015zlo}
C.~Caprini {\em et~al.}, ``{Science with the space-based interferometer eLISA.
  II: Gravitational waves from cosmological phase transitions},''
  \href{http://dx.doi.org/10.1088/1475-7516/2016/04/001}{{\em JCAP} {\bfseries
  1604} no.~04, (2016) 001},
\href{http://arxiv.org/abs/1512.06239}{{\ttfamily arXiv:1512.06239
  [astro-ph.CO]}}.

\bibitem{Cai:2017cbj}
R.-G. Cai, Z.~Cao, Z.-K. Guo, S.-J. Wang, and T.~Yang, ``{The
  Gravitational-Wave Physics},''
\href{http://arxiv.org/abs/1703.00187}{{\ttfamily arXiv:1703.00187 [gr-qc]}}.

\bibitem{Weir:2017wfa}
D.~J. Weir, ``{Gravitational waves from a first order electroweak phase
  transition: a review},''
\newblock 2017.
\newblock \href{http://arxiv.org/abs/1705.01783}{{\ttfamily arXiv:1705.01783
  [hep-ph]}}.
\newblock
\url{http://inspirehep.net/record/1598112/files/arXiv:1705.01783.pdf}.
\newblock

\bibitem{Huang:2016cjm}
P.~Huang, A.~J. Long, and L.-T. Wang, ``{Probing the Electroweak Phase
  Transition with Higgs Factories and Gravitational Waves},''
  \href{http://dx.doi.org/10.1103/PhysRevD.94.075008}{{\em Phys. Rev.}
  {\bfseries D94} no.~7, (2016) 075008},
\href{http://arxiv.org/abs/1608.06619}{{\ttfamily arXiv:1608.06619 [hep-ph]}}.

\bibitem{Hashino:2016rvx}
K.~Hashino, M.~Kakizaki, S.~Kanemura, and T.~Matsui, ``{Synergy between
  measurements of gravitational waves and the triple-Higgs coupling in probing
  the first-order electroweak phase transition},''
  \href{http://dx.doi.org/10.1103/PhysRevD.94.015005}{{\em Phys. Rev.}
  {\bfseries D94} no.~1, (2016) 015005},
\href{http://arxiv.org/abs/1604.02069}{{\ttfamily arXiv:1604.02069 [hep-ph]}}.

\bibitem{Hashino:2016xoj}
K.~Hashino, M.~Kakizaki, S.~Kanemura, P.~Ko, and T.~Matsui, ``{Gravitational
  waves and Higgs boson couplings for exploring first order phase transition in
  the model with a singlet scalar field},''
  \href{http://dx.doi.org/10.1016/j.physletb.2016.12.052}{{\em Phys. Lett.}
  {\bfseries B766} (2017) 49--54},
\href{http://arxiv.org/abs/1609.00297}{{\ttfamily arXiv:1609.00297 [hep-ph]}}.

\bibitem{Beniwal:2017eik}
A.~Beniwal, M.~Lewicki, J.~D. Wells, M.~White, and A.~G. Williams,
  ``{Gravitational wave, collider and dark matter signals from a scalar singlet
  electroweak baryogenesis},''
\href{http://arxiv.org/abs/1702.06124}{{\ttfamily arXiv:1702.06124 [hep-ph]}}.

\bibitem{Croon:2018erz}
D.~Croon, V.~Sanz, and G.~White, ``{Model Discrimination in Gravitational Wave
  spectra from Dark Phase Transitions},''
\href{http://arxiv.org/abs/1806.02332}{{\ttfamily arXiv:1806.02332 [hep-ph]}}.

\bibitem{Coleman:1973jx}
S.~R. Coleman and E.~J. Weinberg, ``{Radiative Corrections as the Origin of
  Spontaneous Symmetry Breaking},''
\href{http://dx.doi.org/10.1103/PhysRevD.7.1888}{{\em Phys. Rev.} {\bfseries
  D7} (1973) 1888--1910}.

\bibitem{Quiros:1999jp}
M.~Quiros, ``{Finite temperature field theory and phase transitions},'' in {\em
  {Proceedings, Summer School in High-energy physics and cosmology: Trieste,
  Italy, June 29-July 17, 1998}}, pp.~187--259.
\newblock 1999.
\newblock
\href{http://arxiv.org/abs/hep-ph/9901312}{{\ttfamily arXiv:hep-ph/9901312
  [hep-ph]}}.
\newblock

\bibitem{Parwani:1991gq}
R.~R. Parwani, ``{Resummation in a hot scalar field theory},''
  \href{http://dx.doi.org/10.1103/PhysRevD.45.4695,
  10.1103/PhysRevD.48.5965.2}{{\em Phys. Rev.} {\bfseries D45} (1992) 4695},
  \href{http://arxiv.org/abs/hep-ph/9204216}{{\ttfamily arXiv:hep-ph/9204216
  [hep-ph]}}.
[Erratum: Phys. Rev.D48,5965(1993)].

\bibitem{Gross:1980br}
D.~J. Gross, R.~D. Pisarski, and L.~G. Yaffe, ``{QCD and Instantons at Finite
  Temperature},''
\href{http://dx.doi.org/10.1103/RevModPhys.53.43}{{\em Rev. Mod. Phys.}
  {\bfseries 53} (1981) 43}.

\bibitem{Nielsen:1975fs}
N.~K. Nielsen, ``{On the Gauge Dependence of Spontaneous Symmetry Breaking in
  Gauge Theories},''
\href{http://dx.doi.org/10.1016/0550-3213(75)90301-6}{{\em Nucl. Phys.}
  {\bfseries B101} (1975) 173--188}.

\bibitem{Patel:2011th}
H.~H. Patel and M.~J. Ramsey-Musolf, ``{Baryon Washout, Electroweak Phase
  Transition, and Perturbation Theory},''
  \href{http://dx.doi.org/10.1007/JHEP07(2011)029}{{\em JHEP} {\bfseries 07}
  (2011) 029},
\href{http://arxiv.org/abs/1101.4665}{{\ttfamily arXiv:1101.4665 [hep-ph]}}.

\bibitem{Chao:2017vrq}
W.~Chao, H.-K. Guo, and J.~Shu, ``{Gravitational Wave Signals of Electroweak
  Phase Transition Triggered by Dark Matter},''
  \href{http://dx.doi.org/10.1088/1475-7516/2017/09/009}{{\em JCAP} {\bfseries
  1709} no.~09, (2017) 009},
\href{http://arxiv.org/abs/1702.02698}{{\ttfamily arXiv:1702.02698 [hep-ph]}}.

\bibitem{Bian:2017wfv}
L.~Bian, H.-K. Guo, and J.~Shu, ``{Gravitational Waves, baryon asymmetry of the
  universe and electric dipole moment in the CP-violating NMSSM},''
\href{http://arxiv.org/abs/1704.02488}{{\ttfamily arXiv:1704.02488 [hep-ph]}}.

\bibitem{Chao:2017ilw}
W.~Chao, W.-F. Cui, H.-K. Guo, and J.~Shu, ``{Gravitational Wave Imprint of New
  Symmetry Breaking},''
\href{http://arxiv.org/abs/1707.09759}{{\ttfamily arXiv:1707.09759 [hep-ph]}}.

\bibitem{Wainwright:2011kj}
C.~L. Wainwright, ``{CosmoTransitions: Computing Cosmological Phase Transition
  Temperatures and Bubble Profiles with Multiple Fields},''
  \href{http://dx.doi.org/10.1016/j.cpc.2012.04.004}{{\em Comput. Phys.
  Commun.} {\bfseries 183} (2012) 2006--2013},
\href{http://arxiv.org/abs/1109.4189}{{\ttfamily arXiv:1109.4189 [hep-ph]}}.

\bibitem{Cline:2006ts}
J.~M. Cline, ``{Baryogenesis},'' in {\em {Les Houches Summer School - Session
  86: Particle Physics and Cosmology: The Fabric of Spacetime Les Houches,
  France, July 31-August 25, 2006}}.
\newblock 2006.
\newblock
\href{http://arxiv.org/abs/hep-ph/0609145}{{\ttfamily arXiv:hep-ph/0609145
  [hep-ph]}}.
\newblock

\bibitem{KurkiSuonio:1995pp}
H.~Kurki-Suonio and M.~Laine, ``{Supersonic deflagrations in cosmological phase
  transitions},'' \href{http://dx.doi.org/10.1103/PhysRevD.51.5431}{{\em Phys.
  Rev.} {\bfseries D51} (1995) 5431--5437},
\href{http://arxiv.org/abs/hep-ph/9501216}{{\ttfamily arXiv:hep-ph/9501216
  [hep-ph]}}.

\bibitem{Steinhardt:1981ct}
P.~J. Steinhardt, ``{Relativistic Detonation Waves and Bubble Growth in False
  Vacuum Decay},''
\href{http://dx.doi.org/10.1103/PhysRevD.25.2074}{{\em Phys. Rev.} {\bfseries
  D25} (1982) 2074}.

\bibitem{Konstandin:2010dm}
T.~Konstandin and J.~M. No, ``{Hydrodynamic obstruction to bubble expansion},''
  \href{http://dx.doi.org/10.1088/1475-7516/2011/02/008}{{\em JCAP} {\bfseries
  1102} (2011) 008},
\href{http://arxiv.org/abs/1011.3735}{{\ttfamily arXiv:1011.3735 [hep-ph]}}.

\bibitem{John:2000zq}
P.~John and M.~G. Schmidt, ``{Do stops slow down electroweak bubble walls?},''
  \href{http://dx.doi.org/10.1016/S0550-3213(00)00768-9,
  10.1016/S0550-3213(02)01014-3}{{\em Nucl. Phys.} {\bfseries B598} (2001)
  291--305}, \href{http://arxiv.org/abs/hep-ph/0002050}{{\ttfamily
  arXiv:hep-ph/0002050 [hep-ph]}}.
[Erratum: Nucl. Phys.B648,449(2003)].

\bibitem{Cirigliano:2006dg}
V.~Cirigliano, S.~Profumo, and M.~J. Ramsey-Musolf, ``{Baryogenesis, Electric
  Dipole Moments and Dark Matter in the MSSM},''
  \href{http://dx.doi.org/10.1088/1126-6708/2006/07/002}{{\em JHEP} {\bfseries
  07} (2006) 002},
\href{http://arxiv.org/abs/hep-ph/0603246}{{\ttfamily arXiv:hep-ph/0603246
  [hep-ph]}}.

\bibitem{Chung:2009qs}
D.~J.~H. Chung, B.~Garbrecht, M.~Ramsey-Musolf, and S.~Tulin, ``{Supergauge
  interactions and electroweak baryogenesis},''
  \href{http://dx.doi.org/10.1088/1126-6708/2009/12/067}{{\em JHEP} {\bfseries
  12} (2009) 067},
\href{http://arxiv.org/abs/0908.2187}{{\ttfamily arXiv:0908.2187 [hep-ph]}}.

\bibitem{Chao:2014dpa}
W.~Chao and M.~J. Ramsey-Musolf, ``{Electroweak Baryogenesis, Electric Dipole
  Moments, and Higgs Diphoton Decays},''
  \href{http://dx.doi.org/10.1007/JHEP10(2014)180}{{\em JHEP} {\bfseries 10}
  (2014) 180},
\href{http://arxiv.org/abs/1406.0517}{{\ttfamily arXiv:1406.0517 [hep-ph]}}.

\bibitem{Guo:2016ixx}
H.-K. Guo, Y.-Y. Li, T.~Liu, M.~Ramsey-Musolf, and J.~Shu, ``{Lepton-Flavored
  Electroweak Baryogenesis},''
  \href{http://dx.doi.org/10.1103/PhysRevD.96.115034}{{\em Phys. Rev.}
  {\bfseries D96} no.~11, (2017) 115034},
\href{http://arxiv.org/abs/1609.09849}{{\ttfamily arXiv:1609.09849 [hep-ph]}}.

\bibitem{White:2016nbo}
G.~A. White, \href{http://dx.doi.org/10.1088/978-1-6817-4457-5}{{\em {A
  Pedagogical Introduction to Electroweak Baryogenesis}}}.
\newblock IOP Concise Physics. Morgan \& Claypool,
2016.
\newblock

\bibitem{Kozaczuk:2015owa}
J.~Kozaczuk, ``{Bubble Expansion and the Viability of Singlet-Driven
  Electroweak Baryogenesis},''
  \href{http://dx.doi.org/10.1007/JHEP10(2015)135}{{\em JHEP} {\bfseries 10}
  (2015) 135},
\href{http://arxiv.org/abs/1506.04741}{{\ttfamily arXiv:1506.04741 [hep-ph]}}.

\bibitem{Kosowsky:1991ua}
A.~Kosowsky, M.~S. Turner, and R.~Watkins, ``{Gravitational radiation from
  colliding vacuum bubbles},''
\href{http://dx.doi.org/10.1103/PhysRevD.45.4514}{{\em Phys. Rev.} {\bfseries
  D45} (1992) 4514--4535}.

\bibitem{Kosowsky:1992rz}
A.~Kosowsky, M.~S. Turner, and R.~Watkins, ``{Gravitational waves from first
  order cosmological phase transitions},''
\href{http://dx.doi.org/10.1103/PhysRevLett.69.2026}{{\em Phys. Rev. Lett.}
  {\bfseries 69} (1992) 2026--2029}.

\bibitem{Kosowsky:1992vn}
A.~Kosowsky and M.~S. Turner, ``{Gravitational radiation from colliding vacuum
  bubbles: envelope approximation to many bubble collisions},''
  \href{http://dx.doi.org/10.1103/PhysRevD.47.4372}{{\em Phys. Rev.} {\bfseries
  D47} (1993) 4372--4391},
\href{http://arxiv.org/abs/astro-ph/9211004}{{\ttfamily arXiv:astro-ph/9211004
  [astro-ph]}}.

\bibitem{Huber:2008hg}
S.~J. Huber and T.~Konstandin, ``{Gravitational Wave Production by Collisions:
  More Bubbles},'' \href{http://dx.doi.org/10.1088/1475-7516/2008/09/022}{{\em
  JCAP} {\bfseries 0809} (2008) 022},
\href{http://arxiv.org/abs/0806.1828}{{\ttfamily arXiv:0806.1828 [hep-ph]}}.

\bibitem{Jinno:2016vai}
R.~Jinno and M.~Takimoto, ``{Gravitational waves from bubble collisions: An
  analytic derivation},''
  \href{http://dx.doi.org/10.1103/PhysRevD.95.024009}{{\em Phys. Rev.}
  {\bfseries D95} no.~2, (2017) 024009},
\href{http://arxiv.org/abs/1605.01403}{{\ttfamily arXiv:1605.01403
  [astro-ph.CO]}}.

\bibitem{Jinno:2017fby}
R.~Jinno and M.~Takimoto, ``{Gravitational waves from bubble dynamics: Beyond
  the Envelope},''
\href{http://arxiv.org/abs/1707.03111}{{\ttfamily arXiv:1707.03111 [hep-ph]}}.

\bibitem{Hindmarsh:2013xza}
M.~Hindmarsh, S.~J. Huber, K.~Rummukainen, and D.~J. Weir, ``{Gravitational
  waves from the sound of a first order phase transition},''
  \href{http://dx.doi.org/10.1103/PhysRevLett.112.041301}{{\em Phys. Rev.
  Lett.} {\bfseries 112} (2014) 041301},
\href{http://arxiv.org/abs/1304.2433}{{\ttfamily arXiv:1304.2433 [hep-ph]}}.

\bibitem{Hindmarsh:2015qta}
M.~Hindmarsh, S.~J. Huber, K.~Rummukainen, and D.~J. Weir, ``{Numerical
  simulations of acoustically generated gravitational waves at a first order
  phase transition},'' \href{http://dx.doi.org/10.1103/PhysRevD.92.123009}{{\em
  Phys. Rev.} {\bfseries D92} no.~12, (2015) 123009},
\href{http://arxiv.org/abs/1504.03291}{{\ttfamily arXiv:1504.03291
  [astro-ph.CO]}}.

\bibitem{Caprini:2009yp}
C.~Caprini, R.~Durrer, and G.~Servant, ``{The stochastic gravitational wave
  background from turbulence and magnetic fields generated by a first-order
  phase transition},''
  \href{http://dx.doi.org/10.1088/1475-7516/2009/12/024}{{\em JCAP} {\bfseries
  0912} (2009) 024},
\href{http://arxiv.org/abs/0909.0622}{{\ttfamily arXiv:0909.0622
  [astro-ph.CO]}}.

\bibitem{Binetruy:2012ze}
P.~Binetruy, A.~Bohe, C.~Caprini, and J.-F. Dufaux, ``{Cosmological Backgrounds
  of Gravitational Waves and eLISA/NGO: Phase Transitions, Cosmic Strings and
  Other Sources},'' \href{http://dx.doi.org/10.1088/1475-7516/2012/06/027}{{\em
  JCAP} {\bfseries 1206} (2012) 027},
\href{http://arxiv.org/abs/1201.0983}{{\ttfamily arXiv:1201.0983 [gr-qc]}}.

\bibitem{Bodeker:2017cim}
D.~Bodeker and G.~D. Moore, ``{Electroweak Bubble Wall Speed Limit},''
  \href{http://dx.doi.org/10.1088/1475-7516/2017/05/025}{{\em JCAP} {\bfseries
  1705} no.~05, (2017) 025},
\href{http://arxiv.org/abs/1703.08215}{{\ttfamily arXiv:1703.08215 [hep-ph]}}.

\bibitem{Hindmarsh:2017gnf}
M.~Hindmarsh, S.~J. Huber, K.~Rummukainen, and D.~J. Weir, ``{Shape of the
  acoustic gravitational wave power spectrum from a first order phase
  transition},''
\href{http://arxiv.org/abs/1704.05871}{{\ttfamily arXiv:1704.05871
  [astro-ph.CO]}}.

\bibitem{Hindmarsh:2016lnk}
M.~Hindmarsh, ``{Sound shell model for acoustic gravitational wave production
  at a first-order phase transition in the early Universe},''
  \href{http://dx.doi.org/10.1103/PhysRevLett.120.071301}{{\em Phys. Rev.
  Lett.} {\bfseries 120} no.~7, (2018) 071301},
\href{http://arxiv.org/abs/1608.04735}{{\ttfamily arXiv:1608.04735
  [astro-ph.CO]}}.

\bibitem{Kahniashvili:2008pe}
T.~Kahniashvili, L.~Campanelli, G.~Gogoberidze, Y.~Maravin, and B.~Ratra,
  ``{Gravitational Radiation from Primordial Helical Inverse Cascade MHD
  Turbulence},'' \href{http://dx.doi.org/10.1103/PhysRevD.78.123006,
  10.1103/PhysRevD.79.109901}{{\em Phys. Rev.} {\bfseries D78} (2008) 123006},
  \href{http://arxiv.org/abs/0809.1899}{{\ttfamily arXiv:0809.1899
  [astro-ph]}}.
[Erratum: Phys. Rev.D79,109901(2009)].

\bibitem{Gong:2014mca}
X.~Gong {\em et~al.}, ``{Descope of the ALIA mission},''
  \href{http://dx.doi.org/10.1088/1742-6596/610/1/012011}{{\em J. Phys. Conf.
  Ser.} {\bfseries 610} no.~1, (2015) 012011},
\href{http://arxiv.org/abs/1410.7296}{{\ttfamily arXiv:1410.7296 [gr-qc]}}.

\bibitem{Luo:2015ght}
{\bfseries TianQin} Collaboration, J.~Luo {\em et~al.}, ``{TianQin: a
  space-borne gravitational wave detector},''
  \href{http://dx.doi.org/10.1088/0264-9381/33/3/035010}{{\em Class. Quant.
  Grav.} {\bfseries 33} no.~3, (2016) 035010},
\href{http://arxiv.org/abs/1512.02076}{{\ttfamily arXiv:1512.02076
  [astro-ph.IM]}}.

\bibitem{Kudoh:2005as}
H.~Kudoh, A.~Taruya, T.~Hiramatsu, and Y.~Himemoto, ``{Detecting a
  gravitational-wave background with next-generation space interferometers},''
  \href{http://dx.doi.org/10.1103/PhysRevD.73.064006}{{\em Phys. Rev.}
  {\bfseries D73} (2006) 064006},
\href{http://arxiv.org/abs/gr-qc/0511145}{{\ttfamily arXiv:gr-qc/0511145
  [gr-qc]}}.

\bibitem{Klein:2015hvg}
A.~Klein {\em et~al.}, ``{Science with the space-based interferometer eLISA:
  Supermassive black hole binaries},''
  \href{http://dx.doi.org/10.1103/PhysRevD.93.024003}{{\em Phys. Rev.}
  {\bfseries D93} no.~2, (2016) 024003},
\href{http://arxiv.org/abs/1511.05581}{{\ttfamily arXiv:1511.05581 [gr-qc]}}.

\bibitem{Thrane:2013oya}
E.~Thrane and J.~D. Romano, ``{Sensitivity curves for searches for
  gravitational-wave backgrounds},''
  \href{http://dx.doi.org/10.1103/PhysRevD.88.124032}{{\em Phys. Rev.}
  {\bfseries D88} no.~12, (2013) 124032},
\href{http://arxiv.org/abs/1310.5300}{{\ttfamily arXiv:1310.5300
  [astro-ph.IM]}}.

\bibitem{Koffas:2017cam}
{\bfseries ATLAS} Collaboration, T.~Koffas, ``{ATLAS Higgs physics prospects at
  the high luminosity LHC},''
{\em PoS} {\bfseries ICHEP2016} (2017) 426.

\bibitem{Davey:2017obx}
{\bfseries ATLAS} Collaboration, W.~Davey, ``{Search for di-Higgs production
  with the ATLAS detector},''
\href{http://dx.doi.org/10.22323/1.314.0272}{{\em PoS} {\bfseries EPS-HEP2017}
  (2017) 272}.

\bibitem{Morse:2017efg}
{\bfseries CMS} Collaboration, D.~M. Morse, ``{Latest results on di-Higgs boson
  production with CMS},''
\newblock 2017.
\newblock \href{http://arxiv.org/abs/1708.08249}{{\ttfamily arXiv:1708.08249
  [hep-ex]}}.
\newblock
\url{http://inspirehep.net/record/1620209/files/arXiv:1708.08249.pdf}.
\newblock

\bibitem{Goncalves:2018qas}
D.~Gonçalves, T.~Han, F.~Kling, T.~Plehn, and M.~Takeuchi, ``{Higgs boson pair
  production at future hadron colliders: From kinematics to dynamics},''
  \href{http://dx.doi.org/10.1103/PhysRevD.97.113004}{{\em Phys. Rev.}
  {\bfseries D97} no.~11, (2018) 113004},
\href{http://arxiv.org/abs/1802.04319}{{\ttfamily arXiv:1802.04319 [hep-ph]}}.

\bibitem{Kim:2018uty}
J.~H. Kim, Y.~Sakaki, and M.~Son, ``{Combined analysis of double Higgs
  production via gluon fusion at the HL-LHC in the effective field theory
  approach},'' \href{http://dx.doi.org/10.1103/PhysRevD.98.015016}{{\em Phys.
  Rev.} {\bfseries D98} no.~1, (2018) 015016},
\href{http://arxiv.org/abs/1801.06093}{{\ttfamily arXiv:1801.06093 [hep-ph]}}.

\bibitem{Kim:2018cxf}
J.~H. Kim, K.~Kong, K.~T. Matchev, and M.~Park, ``{Measuring the Triple Higgs
  Self-Interaction at the Large Hadron Collider},''
\href{http://arxiv.org/abs/1807.11498}{{\ttfamily arXiv:1807.11498 [hep-ph]}}.

\bibitem{Baur:2003gp}
U.~Baur, T.~Plehn, and D.~L. Rainwater, ``{Probing the Higgs selfcoupling at
  hadron colliders using rare decays},''
  \href{http://dx.doi.org/10.1103/PhysRevD.69.053004}{{\em Phys. Rev.}
  {\bfseries D69} (2004) 053004},
\href{http://arxiv.org/abs/hep-ph/0310056}{{\ttfamily arXiv:hep-ph/0310056
  [hep-ph]}}.

\bibitem{Baglio:2012np}
J.~Baglio, A.~Djouadi, R.~Gröber, M.~M. Mühlleitner, J.~Quevillon, and
  M.~Spira, ``{The measurement of the Higgs self-coupling at the LHC:
  theoretical status},'' \href{http://dx.doi.org/10.1007/JHEP04(2013)151}{{\em
  JHEP} {\bfseries 04} (2013) 151},
\href{http://arxiv.org/abs/1212.5581}{{\ttfamily arXiv:1212.5581 [hep-ph]}}.

\bibitem{Huang:2015tdv}
P.~Huang, A.~Joglekar, B.~Li, and C.~E.~M. Wagner, ``{Probing the Electroweak
  Phase Transition at the LHC},''
  \href{http://dx.doi.org/10.1103/PhysRevD.93.055049}{{\em Phys. Rev.}
  {\bfseries D93} no.~5, (2016) 055049},
\href{http://arxiv.org/abs/1512.00068}{{\ttfamily arXiv:1512.00068 [hep-ph]}}.

\bibitem{Azatov:2015oxa}
A.~Azatov, R.~Contino, G.~Panico, and M.~Son, ``{Effective field theory
  analysis of double Higgs boson production via gluon fusion},''
  \href{http://dx.doi.org/10.1103/PhysRevD.92.035001}{{\em Phys. Rev.}
  {\bfseries D92} no.~3, (2015) 035001},
\href{http://arxiv.org/abs/1502.00539}{{\ttfamily arXiv:1502.00539 [hep-ph]}}.

\bibitem{Chang:2018uwu}
J.~Chang, K.~Cheung, J.~S. Lee, C.-T. Lu, and J.~Park, ``{Higgs-boson-pair
  production $H(\rightarrow b\overline{b})H(\rightarrow\gamma\gamma)$ from
  gluon fusion at the HL-LHC and HL-100 TeV hadron collider},''
\href{http://arxiv.org/abs/1804.07130}{{\ttfamily arXiv:1804.07130 [hep-ph]}}.

\bibitem{Ellis:2018gqa}
J.~Ellis, C.~W. Murphy, V.~Sanz, and T.~You, ``{Updated Global SMEFT Fit to
  Higgs, Diboson and Electroweak Data},''
  \href{http://dx.doi.org/10.1007/JHEP06(2018)146}{{\em JHEP} {\bfseries 06}
  (2018) 146},
\href{http://arxiv.org/abs/1803.03252}{{\ttfamily arXiv:1803.03252 [hep-ph]}}.

\bibitem{Baur:2003gpa}
U.~Baur, T.~Plehn, and D.~L. Rainwater, ``{Examining the Higgs boson potential
  at lepton and hadron colliders: A Comparative analysis},''
  \href{http://dx.doi.org/10.1103/PhysRevD.68.033001}{{\em Phys. Rev.}
  {\bfseries D68} (2003) 033001},
\href{http://arxiv.org/abs/hep-ph/0304015}{{\ttfamily arXiv:hep-ph/0304015
  [hep-ph]}}.

\bibitem{Dolan:2012rv}
M.~J. Dolan, C.~Englert, and M.~Spannowsky, ``{Higgs self-coupling measurements
  at the LHC},'' \href{http://dx.doi.org/10.1007/JHEP10(2012)112}{{\em JHEP}
  {\bfseries 10} (2012) 112},
\href{http://arxiv.org/abs/1206.5001}{{\ttfamily arXiv:1206.5001 [hep-ph]}}.

\bibitem{Papaefstathiou:2012qe}
A.~Papaefstathiou, L.~L. Yang, and J.~Zurita, ``{Higgs boson pair production at
  the LHC in the $b \bar{b}W^+ W^-$ channel},''
  \href{http://dx.doi.org/10.1103/PhysRevD.87.011301}{{\em Phys. Rev.}
  {\bfseries D87} no.~1, (2013) 011301},
\href{http://arxiv.org/abs/1209.1489}{{\ttfamily arXiv:1209.1489 [hep-ph]}}.

\bibitem{deLima:2014dta}
D.~E. Ferreira~de Lima, A.~Papaefstathiou, and M.~Spannowsky, ``{Standard model
  Higgs boson pair production in the ( $ b\overline{b} $ )( $ b\overline{b} $ )
  final state},'' \href{http://dx.doi.org/10.1007/JHEP08(2014)030}{{\em JHEP}
  {\bfseries 08} (2014) 030},
\href{http://arxiv.org/abs/1404.7139}{{\ttfamily arXiv:1404.7139 [hep-ph]}}.

\bibitem{Behr:2015oqq}
J.~K. Behr, D.~Bortoletto, J.~A. Frost, N.~P. Hartland, C.~Issever, and
  J.~Rojo, ``{Boosting Higgs pair production in the $b\bar{b}b\bar{b}$ final
  state with multivariate techniques},''
  \href{http://dx.doi.org/10.1140/epjc/s10052-016-4215-5}{{\em Eur. Phys. J.}
  {\bfseries C76} no.~7, (2016) 386},
\href{http://arxiv.org/abs/1512.08928}{{\ttfamily arXiv:1512.08928 [hep-ph]}}.

\bibitem{CerdaAlberich:2018tkw}
{\bfseries ATLAS} Collaboration, L.~Cerda~Alberich, ``{Search for resonant and
  enhanced non-resonant di-Higgs production in the $ \gamma \gamma b \bar b$
  channel with data at 13 TeV with the ATLAS detector},''
\href{http://dx.doi.org/10.22323/1.314.0687}{{\em PoS} {\bfseries EPS-HEP2017}
  (2018) 687}.

\bibitem{Reichert:2017puo}
M.~Reichert, A.~Eichhorn, H.~Gies, J.~M. Pawlowski, T.~Plehn, and M.~M.
  Scherer, ``{Probing baryogenesis through the Higgs boson self-coupling},''
  \href{http://dx.doi.org/10.1103/PhysRevD.97.075008}{{\em Phys. Rev.}
  {\bfseries D97} no.~7, (2018) 075008},
\href{http://arxiv.org/abs/1711.00019}{{\ttfamily arXiv:1711.00019 [hep-ph]}}.

\bibitem{Adhikary:2017jtu}
A.~Adhikary, S.~Banerjee, R.~K. Barman, B.~Bhattacherjee, and S.~Niyogi,
  ``{Revisiting the non-resonant Higgs pair production at the HL-LHC},''
  \href{http://dx.doi.org/10.1007/JHEP07(2018)116}{{\em JHEP} {\bfseries 07}
  (2018) 116},
\href{http://arxiv.org/abs/1712.05346}{{\ttfamily arXiv:1712.05346 [hep-ph]}}.

\bibitem{Chen:2014ask}
C.-Y. Chen, S.~Dawson, and I.~M. Lewis, ``{Exploring resonant di-Higgs boson
  production in the Higgs singlet model},''
  \href{http://dx.doi.org/10.1103/PhysRevD.91.035015}{{\em Phys. Rev.}
  {\bfseries D91} no.~3, (2015) 035015},
\href{http://arxiv.org/abs/1410.5488}{{\ttfamily arXiv:1410.5488 [hep-ph]}}.

\bibitem{Lewis:2017dme}
I.~M. Lewis and M.~Sullivan, ``{Benchmarks for Double Higgs Production in the
  Singlet Extended Standard Model at the LHC},''
  \href{http://dx.doi.org/10.1103/PhysRevD.96.035037}{{\em Phys. Rev.}
  {\bfseries D96} no.~3, (2017) 035037},
\href{http://arxiv.org/abs/1701.08774}{{\ttfamily arXiv:1701.08774 [hep-ph]}}.

\bibitem{Chen:2017qcz}
C.-Y. Chen, J.~Kozaczuk, and I.~M. Lewis, ``{Non-resonant Collider Signatures
  of a Singlet-Driven Electroweak Phase Transition},''
  \href{http://dx.doi.org/10.1007/JHEP08(2017)096}{{\em JHEP} {\bfseries 08}
  (2017) 096},
\href{http://arxiv.org/abs/1704.05844}{{\ttfamily arXiv:1704.05844 [hep-ph]}}.

\bibitem{Barger:2013jfa}
V.~Barger, L.~L. Everett, C.~B. Jackson, and G.~Shaughnessy, ``{Higgs-Pair
  Production and Measurement of the Triscalar Coupling at LHC(8,14)},''
  \href{http://dx.doi.org/10.1016/j.physletb.2013.12.013}{{\em Phys. Lett.}
  {\bfseries B728} (2014) 433--436},
\href{http://arxiv.org/abs/1311.2931}{{\ttfamily arXiv:1311.2931 [hep-ph]}}.

\bibitem{Dawson:2013bba}
S.~Dawson {\em et~al.}, ``{Working Group Report: Higgs Boson},'' in {\em
  {Proceedings, 2013 Community Summer Study on the Future of U.S. Particle
  Physics: Snowmass on the Mississippi (CSS2013): Minneapolis, MN, USA, July
  29-August 6, 2013}}.
\newblock 2013.
\newblock \href{http://arxiv.org/abs/1310.8361}{{\ttfamily arXiv:1310.8361
  [hep-ex]}}.
\newblock
\url{https://inspirehep.net/record/1262795/files/arXiv:1310.8361.pdf}.
\newblock

\bibitem{Alwall:2014hca}
J.~Alwall, R.~Frederix, S.~Frixione, V.~Hirschi, F.~Maltoni, O.~Mattelaer,
  H.~S. Shao, T.~Stelzer, P.~Torrielli, and M.~Zaro, ``{The automated
  computation of tree-level and next-to-leading order differential cross
  sections, and their matching to parton shower simulations},''
  \href{http://dx.doi.org/10.1007/JHEP07(2014)079}{{\em JHEP} {\bfseries 07}
  (2014) 079},
\href{http://arxiv.org/abs/1405.0301}{{\ttfamily arXiv:1405.0301 [hep-ph]}}.

\bibitem{Ball:2013hta}
{\bfseries NNPDF} Collaboration, R.~D. Ball, V.~Bertone, S.~Carrazza,
  L.~Del~Debbio, S.~Forte, A.~Guffanti, N.~P. Hartland, and J.~Rojo, ``{Parton
  distributions with QED corrections},''
  \href{http://dx.doi.org/10.1016/j.nuclphysb.2013.10.010}{{\em Nucl. Phys.}
  {\bfseries B877} (2013) 290--320},
\href{http://arxiv.org/abs/1308.0598}{{\ttfamily arXiv:1308.0598 [hep-ph]}}.

\bibitem{deFlorian:2013uza}
D.~de~Florian and J.~Mazzitelli, ``{Two-loop virtual corrections to Higgs pair
  production},'' \href{http://dx.doi.org/10.1016/j.physletb.2013.06.046}{{\em
  Phys. Lett.} {\bfseries B724} (2013) 306--309},
\href{http://arxiv.org/abs/1305.5206}{{\ttfamily arXiv:1305.5206 [hep-ph]}}.

\bibitem{Catani:2003zt}
S.~Catani, D.~de~Florian, M.~Grazzini, and P.~Nason, ``{Soft gluon resummation
  for Higgs boson production at hadron colliders},''
  \href{http://dx.doi.org/10.1088/1126-6708/2003/07/028}{{\em JHEP} {\bfseries
  07} (2003) 028},
\href{http://arxiv.org/abs/hep-ph/0306211}{{\ttfamily arXiv:hep-ph/0306211
  [hep-ph]}}.

\bibitem{Sjostrand:2014zea}
T.~Sjöstrand, S.~Ask, J.~R. Christiansen, R.~Corke, N.~Desai, P.~Ilten,
  S.~Mrenna, S.~Prestel, C.~O. Rasmussen, and P.~Z. Skands, ``{An Introduction
  to PYTHIA 8.2},'' \href{http://dx.doi.org/10.1016/j.cpc.2015.01.024}{{\em
  Comput. Phys. Commun.} {\bfseries 191} (2015) 159--177},
\href{http://arxiv.org/abs/1410.3012}{{\ttfamily arXiv:1410.3012 [hep-ph]}}.

\bibitem{Cacciari:2011ma}
M.~Cacciari, G.~P. Salam, and G.~Soyez, ``{FastJet User Manual},''
  \href{http://dx.doi.org/10.1140/epjc/s10052-012-1896-2}{{\em Eur. Phys. J.}
  {\bfseries C72} (2012) 1896},
\href{http://arxiv.org/abs/1111.6097}{{\ttfamily arXiv:1111.6097 [hep-ph]}}.

\bibitem{deFavereau:2013fsa}
{\bfseries DELPHES 3} Collaboration, J.~de~Favereau, C.~Delaere, P.~Demin,
  A.~Giammanco, V.~Lemaître, A.~Mertens, and M.~Selvaggi, ``{DELPHES 3, A
  modular framework for fast simulation of a generic collider experiment},''
  \href{http://dx.doi.org/10.1007/JHEP02(2014)057}{{\em JHEP} {\bfseries 02}
  (2014) 057},
\href{http://arxiv.org/abs/1307.6346}{{\ttfamily arXiv:1307.6346 [hep-ex]}}.

\bibitem{Mangano:2006rw}
M.~L. Mangano, M.~Moretti, F.~Piccinini, and M.~Treccani, ``{Matching matrix
  elements and shower evolution for top-quark production in hadronic
  collisions},'' \href{http://dx.doi.org/10.1088/1126-6708/2007/01/013}{{\em
  JHEP} {\bfseries 01} (2007) 013},
\href{http://arxiv.org/abs/hep-ph/0611129}{{\ttfamily arXiv:hep-ph/0611129
  [hep-ph]}}.

\bibitem{Aaltonen:2010jr}
{\bfseries CDF} Collaboration, T.~Aaltonen {\em et~al.}, ``{Observation of
  Single Top Quark Production and Measurement of |Vtb| with CDF},''
  \href{http://dx.doi.org/10.1103/PhysRevD.82.112005}{{\em Phys. Rev.}
  {\bfseries D82} (2010) 112005},
\href{http://arxiv.org/abs/1004.1181}{{\ttfamily arXiv:1004.1181 [hep-ex]}}.

\bibitem{Baldi:2014kfa}
P.~Baldi, P.~Sadowski, and D.~Whiteson, ``{Searching for Exotic Particles in
  High-Energy Physics with Deep Learning},''
  \href{http://dx.doi.org/10.1038/ncomms5308}{{\em Nature Commun.} {\bfseries
  5} (2014) 4308},
\href{http://arxiv.org/abs/1402.4735}{{\ttfamily arXiv:1402.4735 [hep-ph]}}.

\bibitem{Alves:2016htj}
A.~Alves, ``{Stacking machine learning classifiers to identify Higgs bosons at
  the LHC},'' \href{http://dx.doi.org/10.1088/1748-0221/12/05/T05005}{{\em
  JINST} {\bfseries 12} no.~05, (2017) T05005},
\href{http://arxiv.org/abs/1612.07725}{{\ttfamily arXiv:1612.07725 [hep-ph]}}.

\bibitem{Alves:2015dya}
A.~Alves and K.~Sinha, ``{Searches for Dark Matter at the LHC: A Multivariate
  Analysis in the Mono-$Z$ Channel},''
  \href{http://dx.doi.org/10.1103/PhysRevD.92.115013}{{\em Phys. Rev.}
  {\bfseries D92} no.~11, (2015) 115013},
\href{http://arxiv.org/abs/1507.08294}{{\ttfamily arXiv:1507.08294 [hep-ph]}}.

\bibitem{cutoptimize}
A.~Alves, T.~Ghosh, and K.~Sinha, ``{CutOptimize: A Python package fot
  Cut-and-Count Optimization, to be relesead}.''.

\bibitem{Barr:2005dz}
A.~J. Barr, ``{Measuring slepton spin at the LHC},''
  \href{http://dx.doi.org/10.1088/1126-6708/2006/02/042}{{\em JHEP} {\bfseries
  02} (2006) 042},
\href{http://arxiv.org/abs/hep-ph/0511115}{{\ttfamily arXiv:hep-ph/0511115
  [hep-ph]}}.

\bibitem{Alves:2007xt}
A.~Alves and O.~Eboli, ``{Unravelling the sbottom spin at the CERN LHC},''
  \href{http://dx.doi.org/10.1103/PhysRevD.75.115013}{{\em Phys. Rev.}
  {\bfseries D75} (2007) 115013},
\href{http://arxiv.org/abs/0704.0254}{{\ttfamily arXiv:0704.0254 [hep-ph]}}.

\bibitem{LiMa}
T.~Li and Y.-q. Ma, ``{Analysis methods for results in gamma-ray astronomy},''
  \href{http://dx.doi.org/10.1086/161295}{{\em Astrophys. J.} {\bfseries 272}
  (1983) 317}.

\end{thebibliography}\endgroup
\end{document}